\documentclass[onecolumn,journal]{IEEEtran}
\usepackage{amsmath,amsfonts,algorithmic,algorithm,array,bm,color,cite,graphicx,mathtools,multirow,pdfpages,soul,textcomp,xcolor,stfloats,url,verbatim,}
\usepackage[colorlinks=true, allcolors=blue]{hyperref}
\DeclarePairedDelimiter\floor{\lfloor}{\rfloor}

\usepackage{amsmath}
\DeclareMathOperator*{\argmax}{arg\,max}

\begin{document}
\title{Structural Robustness of Complex Networks:\\ A Survey of {\it A Posteriori} Measures}

\author{Yang~Lou,~Lin~Wang,~and~Guanrong~Chen%
	\thanks{Yang Lou is with the Graduate School of Information Science and Technology, Osaka University, Suita, Osaka 565-0871, Japan, and also with the Department of Computer Science, National Yang Ming Chiao Tung University, Hsinchu 300, Taiwan (e-mail: felix.lou@ieee.org).}
	\thanks{Lin Wang is with the Department of Automation, Shanghai Jiao Tong University, Shanghai 200240, China, and also with the Key Laboratory of System Control and Information Processing, Ministry of Education, Shanghai 200240, China (e-mail: wanglin@sjtu.edu.cn).}%
	\thanks{Guanrong Chen is with the Department of Electrical Engineering, City University of Hong Kong, Hong Kong, China (e-mail:eegchen@cityu.edu.hk).}%
	\thanks{This research was supported in parts by the National Natural Science Foundation of China (Nos. 62002249, 61873167), the Foundation of Key Laboratory of System Control and Information Processing, Ministry of Education, China (No. Scip202103), and the Hong Kong Shun Hing Education and Charity Fund (No. 1886992).}%
	\thanks{Citation: \color{blue}{Y. Lou, L. Wang, and G. Chen, ``Structural Robustness of Complex Networks: A Survey of A Posteriori Measures,'' \textit{IEEE Circuits and Systems Magazine}, Volume 23, Issue 1; doi:10.1109/MCAS.2023.3236659 (2023)}
}
\thanks{ (\textit{Corresponding author: Lin Wang})}
}
\markboth{Journal of \LaTeX\ Class Files,~Vol.~XX, No.~YY, February~2023}%
{Shell \MakeLowercase{\textit{et al.}}: A Sample Article Using IEEEtran.cls for IEEE Journals}
\maketitle
\begin{abstract}
	Network robustness is critical for various industrial and social networks against malicious attacks, which has various meanings in different research contexts and here it refers to the ability of a network to sustain its functionality when a fraction of the network fail to work due to attacks. The rapid development of complex networks research indicates special interest and great concern about the network robustness, which is essential for further analyzing and optimizing network structures towards engineering applications. This comprehensive survey distills the important findings and developments of network robustness research, focusing on the \textit{a posteriori} structural robustness measures for single-layer static networks. Specifically, the \textit{a posteriori} robustness measures are reviewed from four perspectives: 1) network functionality, including connectivity, controllability and communication ability, as well as their extensions; 2) malicious attacks, including conventional and computation-based attack strategies; 3) robustness estimation methods using either analytical approximation or machine learning-based prediction; 4) network robustness optimization. Based on the existing measures, a practical threshold of network destruction is introduced, with the suggestion that network robustness should be measured only before reaching the threshold of destruction. Then, \textit{a posteriori} and \textit{a priori} measures are compared experimentally, revealing the advantages of the \textbf{\textit{a posteriori}} measures. Finally, prospective research directions with respect to \textit{a posteriori} robustness measures are recommended.
\end{abstract}


\section{Introduction}

\IEEEPARstart{N}{etwork} robustness has various meanings in different scenarios for different concerns. In this article, it refers to the ability of a network to sustain its normal functionality when a fraction of the network fail to work due to attacks. Today, malicious attacks widely exist in many engineering and technological facilities and processes, which degrade or even destroy certain network functions, typically through destructing the network structural connectivity thereby disabling the network to continue its functioning. It is therefore crucial to strengthen the network robustness against such attacks and failures\cite{Barabasi2016NS,Newman2010N,Cohen2010Book,Chen2014Book,Callaway2000PRL,Holme2002PRE,Shargel2003PRL,Bashan2013NP,Iyer2013PO,Yuan2015PRE,Dey2019PNAS}.  The study of network robustness generally includes measuring and evaluation, attacking and defending, as well as topological optimization\cite{Ellens2013arXiv,Chan2016DMKD,Freitas2022TKDE}. The concerned damage caused by attacks and failures is typically the degeneration or destruction of network functions, such as connectivity\cite{Schneider2011PNAS,Ellens2013arXiv,Freitas2022TKDE}, controllability\cite{Liu2011N,Yuan2013NC,Xiang2019CSM}, data transmission and communication abilities\cite{Lu2019CL,Mburano2021ISNCC}, and so on.  Among these functions, network connectivity is fundamental and essential to support other functions, although good connectivity does not necessarily guarantee good performance of a certain function on the network. In this regard, the subject of network connectivity, controllability, and communication robustness is of fundamental and practical importance, which has been extensively investigated with applications to, for example, the fields of nervous systems\cite{Yan2017NAT}, wireless sensor networks\cite{Qiu2019TN}, power grids\cite{Chen2017TCASII}, and transportation networks\cite{Yang2018TITS,Cai2020SSCI}, among many others.  This survey article focuses on measuring the network structural robustness with respect to network functions, in particular the network connectivity, controllability and communication ability against destructive attacks. This survey only discusses the robustness of single-layer networks with static connections, since the structural robustness is not the main issue for networks with dynamic and temporal connections.

Measuring is the first step in analyzing and optimizing the network robustness. There are quite many network robustness measures. In this paper, robustness measures are categorized into two classes according to whether attack simulations are needed for the measurement, namely, the \textit{a priori} measures that do not require attack simulations and the \textit{a posteriori} measures that require so.

\textit{A priori} measures are generally quantified by certain indicative network features that can be calculated without performing attack simulations, including: 1) topological measures, e.g., binary connectivity\cite{Diestel2017GT}, efficiency\cite{Latora2007NJP}, betweenness centrality\cite{Freeman1977Soc}, and clustering coefficient\cite{Watts1998N}; 2) adjacency matrix-based spectral measures, e.g., spectral radius\cite{Van2011PRE}, spectral gap\cite{Malliaros2012ICDM}, natural connectivity\cite{Wu2010CPL}; and 3) Laplacian matrix-based spectral measures, e.g., algebraic connectivity\cite{Fiedler1973CMJ}, effective resistance\cite{Klein1993JMC}, and the number of spanning trees\cite{Butler2008Book}. \textit{A priori} measures require only one-time calculation and usually have lower time and computational complexities comparing to \textit{a posteriori} measures\cite{Freitas2022TKDE,Chan2016DMKD}.

\textit{A posteriori} measures, on the other hand, are quantified by the sequence of values that record the concerned functionality of the remaining network after a sequence of node- or edge-attacks, typically removal attacks. The ratios of largest connected components (LCC)\cite{Schneider2011PNAS}, driver nodes (DN)\cite{Liu2011N,Yuan2013NC} and communicable node pairs (CNP)\cite{Lu2019CL,Mburano2021ISNCC} are the most widely-used measures for the connectivity, controllability, and communication ability, respectively. In turn, the robustness of connectivity, controllability, and communication ability is quantified by a sequence of values that record the corresponding remaining measures after a sequence of node- or edge-attacks, respectively. A network is considered to be more robust if it can maintain higher values of the fractions of nodes in LCC and CNP, but lower fractions of DN, throughout the attack process.

\begin{figure*}[htbp]
	\centering
	\includegraphics[width=.8\linewidth]{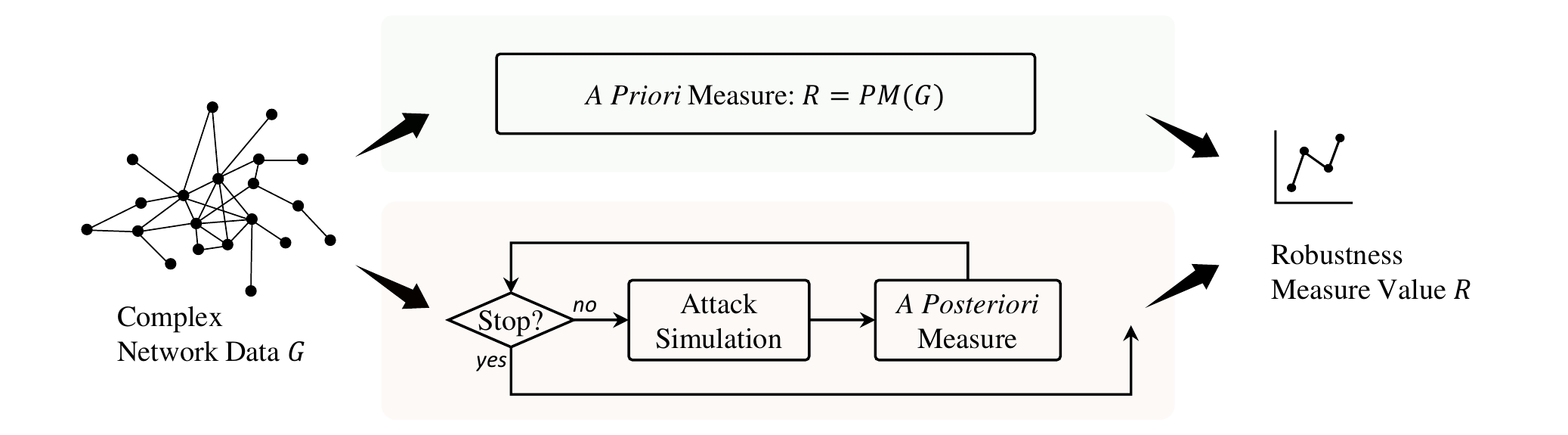}
	\caption{General framework for \textit{a priori} and \textit{a posteriori} measures of network robustness. \textit{A priori} measures perform one-time calculations to evaluate the network robustness, where \textit{PM} represents an \textit{a priori} measure; while \textit{a posteriori} measures require an iterative process to gain the robustness.}\label{fig:mes_cmp}
\end{figure*}

\textit{A priori} measures are easy-to-access and predictive; while \textit{a posteriori} measures are iteratively calculated after each of the sequence of (simulated) attacks, which are usually time-consuming especially for large-scale networks. However, the predictive \textit{a priori} measures have limited scopes of applications\cite{Yamashita2019COMPSAC}. Moreover, the \textit{a posteriori} measures are effective when the attack process is terminated by a specific criterion, whereas the \textit{a priori} measures do not consider the stopping criteria.  Therefore, the time-consuming but precise \textit{a posteriori} measures remain to be the main approach for real-world applications today.

The general framework of \textit{a priori} and \textit{a posteriori} measures is shown in Fig. \ref{fig:mes_cmp}, which shows that \textit{a priori} measures evaluate the network robustness in a straightforward one-time process; while \textit{a posteriori} measures require an iterative process until the stopping criterion is met. It is clear that \textit{a posteriori} measures could have different options on the configuration of stopping criteria and attack strategies, while this is invalid for \textit{a priori} measures.

With desirable robustness measure(s) chosen and used as the objective(s) to optimize, network robustness can be enhanced by model design\cite{Sha2013PCS,Yan2016SR,Lou2018TCASI,Hayashi2018NS,Lou2019R}, edge addition\cite{Beygelzimer2005PA,Freitas2022TKDE}, or edge rewiring\cite{Wu2011PRE,Zeng2012PRE,Louzada2013JCN,Schneider2013SR,Liang2015CPL,Chan2016DMKD,Lou2021TCASII,Wang2020TEVC,Wang2021TEVC}. In so doing, whether or not a modification of the network structure can enhance the robustness has to be evaluated, usually by using \textit{a posteriori} robustness measures that usually requires attack simulations.

Other than attack simulations, network robustness can also be estimated using both analytical and computational methods without iterative calculation. Analytical approximations are applicable when the \textit{a prior} knowledge of the concerned network is available and the attack strategy can be well modeled\cite{Sun2019ICSRS,Sun2021TNSM,Lou2023IJCAS},  e.g., random attacks. In contrast, computational methods are generally data-driven and thus applicable to any attack methods with or without a specific pattern\cite{Lou2022TCYB,Lou2022TNNLS,Lou2023NN}.

In the literature, some general survey papers emphasizing more on \textit{a priori} measures of network robustness are available\cite{Freitas2022TKDE,Ellens2013arXiv}, but there does not seem to be any that specifically emphasizes on the \textit{a posteriori} measures. To fill the gap, this article presents a survey of the \textit{a posteriori} measures of network robustness, including definitions, computation, applications, and optimization. The main contributions of this survey are summarized as follows:

\begin{itemize}
	\item[1)] The \textit{a posteriori} robustness measures are summarized and compared, from the perspectives of network functionality, attack strategies, robustness performance prediction, and structural optimization.
	\item[2)] A threshold of network destruction is proposed, which suggests a more practical robustness measure of the functionality, especially when a network has been severely destructed.
	\item[3)] Both \textit{a posteriori} and \textit{a priori} robustness measures are experimentally compared on a series of directed and undirected network examples. It is found from simulations that \textit{a posteriori} measures have broader applicability.
	\item[4)] Some possible future research directions with respect to network robustness are suggested.
\end{itemize}

The remainder of this article is organized as follows.
Section \ref{sec:pst} reviews the \textit{a posteriori} measures of network robustness, from the perspectives of network functionality, malicious attacks, robustness performance prediction, and optimization.
Section \ref{sec:des} introduces a threshold of network destruction.
Section \ref{sec:cmp} experimentally compares \textit{a posteriori} and \textit{a priori} measures.
Section \ref{sec:ftw} presents some prospective research directions with respect to \textit{a posteriori} robustness measures of complex networks.
Section \ref{sec:end} concludes the survey.

\section{\textit{A Posteriori} Measures for Network Robustness}\label{sec:pst}

Network robustness can be defined differently with different practical meanings in graph theory, control systems, communication networks, biological structures, transportation frameworks, etc.\cite{Barabasi2004NRG,Kitano2004NRG,Boccaletti2006PR,Liu2017FCS,Logins2021TKDE,Logins2020WWW}.
On the one hand, it is possible to consider different robustness measures for the same network. On the other hand, the same measure might be applied to different scenarios, for example to both power grids\cite{Cuadra2015ENG} and food webs\cite{Bellingeri2013TE}, where the main concern is the remaining largest connected components after suffering attacks, and the attacks can be physical or cyber destruction to power stations in power grids or extinction of species in food webs.

Here, the focus is on the ability of a network to sustain its specific function(s) when a fraction of the network fail to work due to attacks. Random failures and malicious attacks\cite{Wandelt2022RESS} occur on nodes or edges, or both, in the form of removal or malfunctioning, under different conditions.  In implementation, the consequence of attacking a node could be either removal or malfunctioning (without removal), while that of attacking an edge is only edge removal in typical cases. When a node is attacked and removed, all of its connected edges will also be removed; while under edge-attacks, no nodes will be removed.

The remainder of this section is organized as follows. Subsection \ref{sub:fun} reviews the \textit{a posteriori} robustness measures from three commonly concerned network functions: connectivity, controllability, and communication ability. Some extensions of these measures will be discussed in Subsection \ref{sub:othmes}. Various attack strategies and robustness estimation methods are reviewed in Subsection \ref{sub:atk} and Subsection \ref{sub:pre}, respectively. Finally, Subsection \ref{sub:opt} presents some robustness optimization technicians based on \textit{a posteriori} measures.

\subsection{Network Functions}\label{sub:fun}

\textit{A posteriori} measures iteratively calculate specifically concerned network function(s) after each occurrence of attacks. The general form of \textit{a posteriori} robustness measures is as follows:
\begin{equation}\label{eq:r}
	R=\frac{1}{K}\sum_{i=1}^{K}{w_i\cdot f(i)}\,,
\end{equation}
where $f(i)$ represents the residual functionality of the remaining network after a number (or proportion) of $i$ objects (either nodes or edges) have been attacked; $K$ represents the total number of attacks; $w_i$ represents the weight of $f(i)$ in calculating $R$. When different network functions are concerned, such as the connectivity, controllability or communication ability, to be discussed below, $f(\cdot)$ will be specified accordingly. The weighting parameter $w$ is usually considered as a parameter for normalization, such that the robustness performances of different-sized networks can be compared. However, these weights also shift the importance of the attacks in the attack sequence, which is often overlooked.

\begin{figure}[htbp]
	\centering
	\includegraphics[width=.5\linewidth]{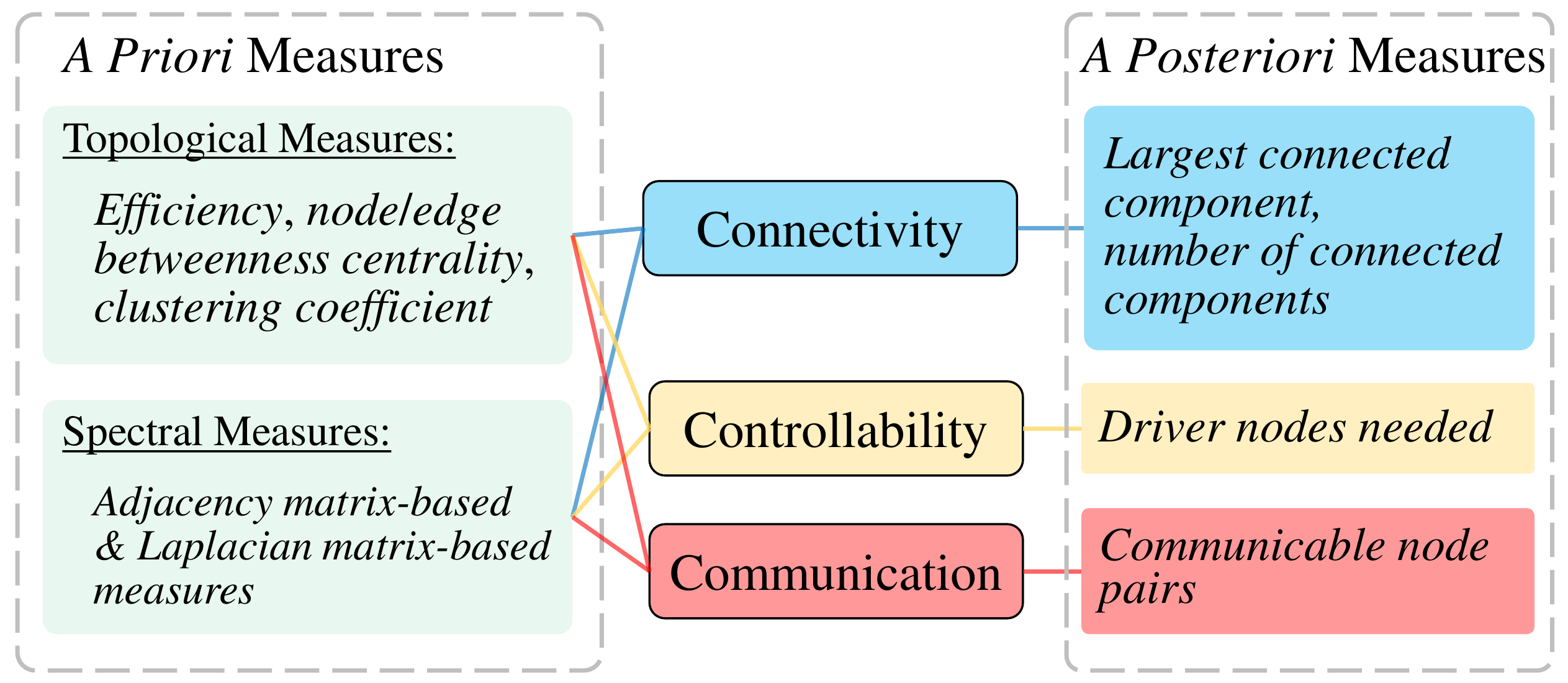}
	\caption{[color online] Widely-used \textit{a priori} and \textit{a posteriori} robustness measures of the three network functions. }\label{fig:functionality}
\end{figure}

Figure \ref{fig:functionality} shows the widely-used \textit{a priori} and \textit{a posteriori} robustness measures of the three network functions. The details of \textit{a posteriori} measures are summarized in the following.

\subsubsection{Connectivity Robustness}

The connectivity of an undirected network means that there is at least one path between any pair of nodes. For a directed network, it is strongly connected if there is at least one directed path from any node to an other node, while it is weakly connected if its underlying undirected network is connected.

LCC is the most commonly-used \textit{a posteriori} measure for connectivity robustness. Under a sequence of node-malfunctioning failures or removals, the connectivity robustness is evaluated by calculating the remaining LCC after each attack\cite{Schneider2011PNAS}, formulated as follows:
\begin{equation}\label{eq:lc}
	R_1=\frac{1}{N}\sum_{i=0}^{N-1}{n_L(i)}=\frac{1}{N}\sum_{i=0}^{N-1}{\frac{N_L(i)}{N}}\,,
\end{equation}
where $n_L(i)$ and $N_L(i)$ represent the proportion and the number of nodes in the remaining LCC after a total number of $i$ nodes have been attacked. Specifically, $f(i)=N_L(i)$ measures the remaining connectivity and $w_i=1/N$ is the uniform weight, assuming that the malfunctioned nodes are still counted as a part of the $N$-node networks.

In contrast, if the attacked nodes are removed from the network, its connectivity robustness is calculated by
\begin{equation}\label{eq:lc2}
	R_2=\frac{1}{N}\sum_{i=0}^{N-1}{\frac{N_L(i)}{N-i}}\,,
\end{equation}
where $w_i=1/(N-i)$ means that the $i$th attacked nodes have been removed from the network. Compared to Eq. (\ref{eq:lc}), $w_i=1/(N-i)$ assigns higher weights to the later attack stages as $i$ increases. Different weighting parameters $w_i$ also change the range of robustness measure, where $R_1\in[0,0.5]$ but $R_2\in[0,1]$.

The measure shown in Eq. (\ref{eq:lc}) for network robustness under node-attacks can be extended to edge-attacks\cite{Zeng2012PRE}, as follows:
\begin{equation}\label{eq:lce}
	R_1^{e}=\frac{1}{M+1}\sum_{i=0}^{M}{n^{e}_L(i)}=\frac{1}{M+1}\sum_{i=0}^{M}{\frac{N^{e}_L(i)}{N}}\,,
\end{equation}
with the superscript $e$ indicating edge-attacks, where the denominator remains $N$ under the assumption that the number of nodes is unchanged during edge-attacks.

When the values $n_L(i)$ or $n^{e}_L(i)$ are plotted, a curve is obtained, which is called the connectivity curve. A higher $R_1$, $R_2$, or $R^{e}_1$ value indicates an overall better connectivity robustness against attacks.

\subsubsection{Controllability Robustness}

Controllability robustness reflects how well a networked system is in maintaining its controllable state. Consider a general linear time-invariant (LTI) networked system, $\dot{{\bf x}}={\bf A}{\bf x}+{\bf B}{\bf u}$, where ${\bf{x}}\in\mathbb{R}^N$ and ${\bf{u}}\in\mathbb{R}^b$ are the state vector and control input, respectively, and ${\bf A}\in\mathbb{R}^{N\times N}$ and ${\bf B}\in\mathbb{R}^{N\times b}$ are constant matrices of compatible dimensions. Conceptually, this LTI system is state controllable if and only if there exists a control input $\bf{u}$ that can drive the system state $\bf{x}$ from any initial state to any target state in the state space within finite time. A commonly-used criterion is that the LTI system is state controllable if and only if the controllability matrix ${\bf C}=[{\bf B}\ {\bf AB}\ {\bf A}^2{\bf B}\ \cdots {\bf A}^{N-1}{\bf B}]$ has a full row-rank\cite{Chen1998Book}. The concept of structural controllability is a slight generalization of the state controllability, to deal with two parameterized matrices ${\bf A}$ and ${\bf B}$, in which the parameters characterize the structure of the underlying system in the sense that if there are specific parameter values ensuring the system to be state controllable then the system is structurally controllable.

When considering a network of many LTI systems, the node system with control input is called a driver node (DN). Network controllability is investigated from two aspects: 1) to gain the full control of the entire dynamical system, one aims to determine how many and which nodes to control\cite{Liu2011N,Yuan2013NC}; 2) for each single node, the aim is to determine the dimension of its controllable subspace\cite{Hosoe1980TAC,Liu2012PO}.

Define the density of DNs by $n_D=\frac{N_D}{N}$, where $N_D$ is the minimum number of DNs needed to retain a full control of the network, which can be calculated using either the minimum inputs theorem (MIT)\cite{Liu2011N} for directed networks or the exact controllability theorem (ECT)\cite{Yuan2013NC} for both directed and undirected networks, defined as follows:
\begin{equation}\label{eq:2nd}
	N_D=
	\begin{cases}
		\text{max}\{1, N-|E^*|\},& \text{using MIT\cite{Liu2011N}},\\
		\text{max}\{1, N-\text{rank}(A)\},& \text{using ECT\cite{Yuan2013NC}},\\
	\end{cases}
\end{equation}
where $|E^*|$ represents the number of edges in the maximum matching $E^*$, which is a basic concept in classical graph theory\cite{Liu2011N}. Under node-attacks, the controllability robustness is measured by
\begin{equation}\label{eq:nd}
	R_3=\frac{1}{N}\sum_{i=0}^{N-1}{n_D(i)}=\frac{1}{N}\sum_{i=0}^{N-1}{\frac{N_D(i)}{N'}}\,,
\end{equation}
where $n_D(i)$ and $N_D(i)$ represent the density and number of DNs needed to retain the network controllability after a total of $i$ nodes have been attacked; $N'$ can be set to either $N'=N-i$ or $N'=N$, depending on specific preference, namely whether or not an attacked node still belongs to the network depends on the situation under consideration. Usually, attacked nodes are assumed to be malfunctioned (but still in the system) in connectivity robustness measures, but will be removed from the network in controllability robustness measures.

Similarly, controllability robustness under edge-attacks is measured by
\begin{equation}\label{eq:nde}
	R^{e}_3=\frac{1}{M+1}\sum_{i=0}^{M}{n^{e}_D(i)}=\frac{1}{M+1}\sum_{i=0}^{M}{\frac{N^{e}_D(i)}{N}}\,,
\end{equation}
where $M$ is the number of edges in the network. When the values $n_D(i)$ or $n^{e}_D(i)$ are plotted, a curve is obtained, which is called the controllability curve. A lower $R_3$ or $R^{e}_3$ value represents a more robust controllability against attacks.

Different from considering the density of DNs, the control centrality measures the control ability of a single node in a directed network\cite{Liu2012PO}, defined by $c^{(j)}_c=C^{(j)}_c/N$, where $C^{(j)}_c=\text{rank}_g({\bf C}^{(j)})$ is the generic dimension of the controllable subspace of node $j$ that can be calculated according to the Hosoe theorem\cite{Hosoe1980TAC}; ${\bf C}$ represents the controllability matrix. Under this measure, the greater the $c^{(j)}_c$ value is, the more ``powerful'' the node $j$ is as a DN.

The expected robust control centrality (ERCC)\cite{Usman2020MSC,Usman2019CCC} is a control centrality-based robustness measure for node-attacks, defined as follows:
\begin{equation}\label{eq:ercc}
	R_{4}^{(j)}(i)=E[C_c^{(j)}(i)]\,,
\end{equation}
where $C_c^{(j)}(i)$ represents the control centrality of node $j$ after a total number of $i$ nodes have been attacked; $E[\cdot]$ is the statistical expectation. The generic robust control centrality (GRCC)\cite{Usman2020MSC,Usman2019CCC} is a generalization of ERCC, defined as follows:
\begin{equation}\label{eq:grcc}
	R_{4}^{e,(j)}(\{e\})=E[C_c^{(j)}(\{e\})]\,,
\end{equation}
where $C_c^{(j)}(\{e\})$ represents the control centrality of node $j$ after a set of edges $\{e\}$ have been removed, under either node- or edge-attacks. Both ERCC and GRCC measure the significance of a single node in controlling part of the system, under random node- and edge-attacks, respectively.

The reachability-based controllability robustness\cite{Parekh2014ICSIT,Sun2021TNSM} is also a control centrality-based robustness measure. Given a fixed number of $H$ controllers that can be pinned anywhere (``free control'' mode), the controllability robustness is calculated by
\begin{equation}\label{eq:freec}
	R_5=\frac{1}{N}\sum_{i=0}^{N-1}{\sum_{j=1}^{H}} {\frac{c_c(j)}{N'}}\,,
\end{equation}
where $\sum_{j=1}^{H}{c_c(j)}$ represents the dimension of the controllable subspace by the given $H$ DNs. During the attack, these DNs can be freely set in the remaining network, as long as the control centrality is maximized. Again, $N'=N-i$ or $N$, depending on the specific situation under consideration.

In the case that the given external controllers are fixedly pinned at a set of given nodes (``fixed control'' mode)\cite{Sun2021TNSM}, the controllability robustness is also measured using Eq. (\ref{eq:freec}), where however $\sum_{j=1}^{H}{c_c(j)}$ counts the dimension of the controllable subspace by the given $H$ fixed controllers.

\subsubsection{Communication Robustness}

Different from the \textit{a priori} measures of general connectivity robustness, which are either spectral measures or topological features, the \textit{a priori} measures of communication robustness are more comprehensive. For example, the $r$-robustness\cite{LeBlanc2013JSAC,Zhao2014CDC,Zhang2015TCNS} based on reachability, and the comprehensive measure proposed in\cite{Morales2018CL} consisting of three indices, including edge betweenness centrality, number of edge cut-sets, and node Wiener impact\cite{Wiener1947JACS}. Nevertheless, the \textit{a posteriori} measures for connectivity robustness remain useful for measuring communication robustness\cite{Qiu2017TN}.

The CNP-based robustness measure is a widely-used \textit{a posteriori} measure for communication robustness, defined as follows\cite{Lu2019CL}:
\begin{equation}\label{eq:com}
	R_6=\frac{1}{N}\sum_{i=0}^{N-1}{\sum_{j=1}^{\Gamma(i)}} {\frac{{S_j\choose 2}}{{N\choose 2}}}\,,
\end{equation}
where $\Gamma(i)$ represents the number of connected components in the remaining network after a total of $i$ nodes have been attacked; $S_j$ represents the number of nodes in the $j$th connected component; ${S_j\choose 2}$ represents the number of communicable node pairs, while ${N\choose 2}$ is the number of all possible node pairs. When ${S_j\choose 2}={N\choose 2}$, the network is fully connected, thus each pair of nodes are communicable; while for the networks that are not fully connected, the number of communicable node pairs should be less than the all possible node pairs, namely ${S_j\choose 2}<{N\choose 2}$.

The following simplified communication robustness\cite{Mburano2021ISNCC} provides a simpler CNP-based measure:
\begin{equation}\label{eq:scom}
	R_7	=\frac{1}{N}\sum_{i=0}^{N-1}{\sum_{j=1}^{\Gamma(i)}} {\frac{{S_j^2}}{{N^2}}} \,,
\end{equation}
which ignores the non-dominant terms in Eq. (\ref{eq:com}) but keeps only the dominant ones.  The computation complexities of both measures are the same, $O(NM)$.

When the CNP values are plotted, a curve is obtained, which is called the communication curve. Apparently, higher values of $R_6$ or $R_7$ represent better communication robustness against attacks.

\subsection{Variants of Robustness Measures}\label{sub:othmes}

Based on the fundamental \textit{a posteriori} robustness measures presented in Subsection \ref{sub:fun}, several variants have been developed with different concerns.

\subsubsection{Rank-based Measure}

Before being attacked, the initial proportions of LCC for all connected networks are the same, namely $n_L(0)=1$. In contrast, the initially required proportion of DNs to fully control a network varies from case to case. This inequality of initial controllability may influence the measurement of robustness.The rank-based controllability measure offers an alternative to diminish this influence, which is defined by
\begin{equation}\label{eq:rank}
	R_8=\frac{1}{N}\sum_{i=0}^{N-1}{r_D(i)}\,,
\end{equation}
where $r_D(i)$ is the rank of the controllability matrix after a total of $i$ nodes have been attacked. Lower ranks are assigned to the networks that possess better controllability.

\begin{figure}[htbp]
	\centering
	\includegraphics[width=0.35\linewidth]{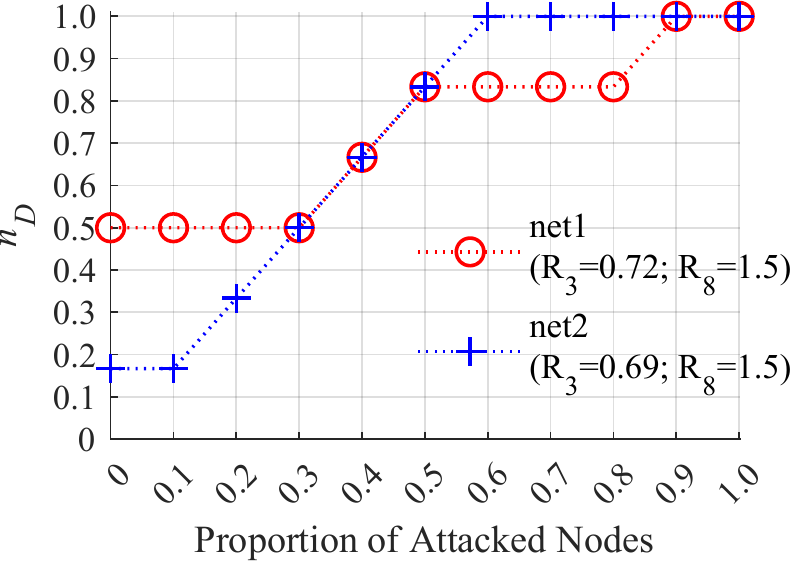}
	\caption{[color online] Example of two different controllability robustness measures. $R_3$ and $R_8$ are calculated using Eqs. (\ref{eq:nd}) and (\ref{eq:rank}), respectively. }\label{fig:eg_rank_measure}
\end{figure}

Figure \ref{fig:eg_rank_measure} shows an illustrative example, where net1 requires a larger initial proportion of DNs than net2. The controllability curve of net1 is flatter than that of net2. Under two different measures, $R_3$ returns that net2 has better controllability robustness than net1, but $R_8$ returns that they have same performance. Clearly, $R_8$ diminishes the influence of the initial states.

\subsubsection{Combinatorial Measure}

Although connectivity robustness has a certain positive correlation with controllability robustness and communication robustness, they actually have very different measures and objectives. In general, good connectivity is the prerequisite for good controllability and communication ability, but the former does not guarantee the latter in general.

Considering connectivity robustness and controllability robustness together, adjustment is necessary since better robustness means maximization Eq. (\ref{eq:lc}) but minimization Eq. (\ref{eq:nd}). To unify them (e.g., both being maximization), a combinatorial measure can be defined using either the opposite of $n_D(i)$\cite{Xiao2014CPB}, as follows:
\begin{equation}\label{eq:xiao}
	R_{9}=\frac{1}{N}\sum_{i=0}^{N-1}(1-n_D(i)) \,,
\end{equation}
or the reciprocal of $n_D(i)$\cite{Wang2018TNSE}, as follows:
\begin{equation}\label{eq:wang_tnse}
	R_{10}	=\frac{1}{N}\sum_{i=0}^{N-1}\frac{n_L(i)}{n_D(i)}\,.
\end{equation}
Maximizing $R_9$ is equivalent to minimizing $R_3$, while maximizing $R_{10}$ is equivalent to either maximizing $R_1$, or minimizing $R_3$, or maximizing $R_1$ and minimizing $R_3$ together.

\subsubsection{Averaged Measure}

All the above-mentioned \textit{a posteriori} robustness measures, except for ERCC and GRCC\cite{Usman2020MSC,Usman2019CCC}, are calculated based on a specific attack sequence, namely, each robustness value is one-to-one corresponding to a specific attack sequence. If network robustness is required to be measured by a number of repeated simulations, or several different attack sequences are required to be considered, then the averaged robustness measure can be applied, which is defined as follows:
\begin{equation}\label{eq:rpt}
	R_{11}=\frac{1}{P\cdot Q}\sum_{p=1}^{P}\sum_{q=1}^{Q}R_{p,q}\,,
\end{equation}
where $R_{p,q}$ represents the network robustness measured under the $p$-th repeated simulation using the $q$-th attack strategy; $P$ is the number of repeated attack simulations; $Q$ is the number of different attack strategies. After averaging, a robustness value will not be corresponding to a specific attack strategy or sequence.

\subsubsection{Other Measures}

When cascading failure-based attacks are considered, the robustness measure can be slightly modified, as follows:
\begin{equation}\label{eq:tang}
	R_{12}=\frac{1}{N}\sum_{h=1}^{H}f(h)\,,
\end{equation}
where $H$ is the required number of attacks to achieve the attack task, for example, a significant destruction of functionality\cite{Pu2012PA,Nie2014PO,Tang2016SR,Chen2017PA,Hou2019ICISBDE}. Here, $H\leq N$ implies that it is not always necessary to attack all nodes in order to destroy the network functions.

When the community structure is concerned, the community robustness can also be calculated using Eq. (\ref{eq:r}), where $f(i)$ could be either the community integrity that counts the number of remaining nodes in the community\cite{Ma2013PRE}, or the normalized mutual information\cite{Wang2017JSTAT}.

It is noted that this survey paper focuses on reviewing the robustness measures of the networks with static connection, whereas the networks with dynamic and temporal connections are not discussed. This is because the robustness measures of dynamic and temporal networks have very different characteristics and applications. For example, the robustness of a temporal network is measured by calculating the relative loss of efficiency caused by attacks\cite{Scellato2011TMC}, as follows:
\begin{equation}\label{eq:temporal}
	R_{13}=1-\frac{\Delta{\epsilon}}{\epsilon_0}\,,
\end{equation}
where $\epsilon_0$ represents the global efficiency of the temporal network within a given time window, and $\Delta{\epsilon}$ represents the efficiency loss caused by attacks. Although it may be regarded as an \textit{a posteriori} measure, it has a different form from Eq. (\ref{eq:r}) that performs iterative attack-and-evaluation operations.

\subsection{Attack Strategies}\label{sub:atk}

From the attacker's perspective, searching for the most destructive attacking sequence is a desirable task, which can also help the defender in considering how to design a best possible network topology with the strongest robustness. Therefore, attack strategy is also a focal topic in the study of network robustness.

For a given network, \textit{a priori} measures return a single deterministic value about the network robustness, which will not change when different attack strategies or different numbers (rounds) of attacks are applied. In contrast, \textit{a posteriori} measures are able to reflect different robustness performances when attack strategies (or attack sequences) vary. The issue of network robustness within different contexts has been extensively investigated, with many edge- and node-attack strategies proposed to destruct the network functions, regarding the connectivity, controllability, communication ability, and so on.

Random attacks remove or malfunction randomly-selected objects (nodes or edges), while targeted attacks aim at attacking deliberately-selected objects, for example, the highest-degree node or the largest-betweenness edge. Given an importance measure $g$ for either nodes or edges, targeted attacks perform sequential attacks to object $j$, with $\argmax g$, meaning that object $j$ is the most important according to measure $g$.

\subsubsection{Degree- and Betweenness-based Attack Strategies}\label{sub:dbc}

For targeted attacks, it is assumed that the targeted object is more important than the others in maintaining the network functionality.  The most frequently-used measures of importance are the degree centrality and betweenness centrality, for both nodes and edges. In fact, the maximum degree-based targeted attack (MDTA) and maximum betweenness-based targeted attack (MBTA) are the most widely-used strategies.

To integrate multiple importance measures into one, weights and probabilities may be considered:
\begin{equation}\label{eq:pj}
	p_j=\sum_i \alpha_i\times\frac{g_{i,j}}{\sum_{j=1}^{K}{g_{i,j}}}\,,
\end{equation}
where $p_j$ is the probability of attacking object $j$; $\alpha_i$ is the weight for importance measure $g_i$; $g_{i,j}$ is the importance measure $g_i$ for object $j$. For example, $p_j=\alpha_1\times\frac{k_j}{\sum_{j=1}^{K}{k_j}} + \alpha_2\times\frac{b_j}{\sum_{j=1}^{K}{b_j}}$ represents a combination of degree and betweenness, where $k_j$ and $b_j$ are the degree and the betweenness of node $j$; weights $\alpha_1$ and $\alpha_2$ adjust the distributions of different features, which can be set manually\cite{Nie2015PA}, or with $\alpha_2$ being replaced by $1-\alpha_1$\cite{Gao2018PA}.

Similarly, three parameters can be used\cite{Hao2020PA} to control the weights of degree, betweenness and harmonic closeness, respectively. Attacking the highest-betweenness node inside the LCC makes MBTA more destructive in the later stages of the attack process\cite{Nguyen2019PA}. These measures have also been used in some strategies to attack interdependent networks\cite{Huang2011PRE,Dong2012PRE,Gao2018PA,Cui2018PA,Hao2020PA}, networks of networks\cite{Dong2013PRE,Liu2015CSF}, and weighted networks\cite{Bellingeri2018PA}.

Both MDTA and MBTA are not only destructive to connectivity robustness, but also effectively degrade other network functions such as controllability and communication ability\cite{Pu2012PA,Nie2014PO,Chen2019TCASII,Lu2019CL}.

\subsubsection{Topology-based Attack Strategies}

Beside degree and betweenness, commonly-used measures of importance include closeness\cite{Borgatti2005SN}, Katz centrality\cite{Katz1953PSY}, neighborhood similarity\cite{Ruan2017APS}, branch weighting\cite{Simon2017MOE}, structural holes\cite{Yang2020SYM}, and so on. However, ranking the importance of nodes or edges is practically intractable for large-scale networks, since most measures cannot guarantee that removing the targeted object will globally and consistently cause the greatest damage to the network.

The hierarchical structure of a directed network enables the random upstream (or downstream) attack to the network controllability, which results in a more destructive attack strategy than random attacks\cite{Liu2012PO}. The module-based attack strategy\cite{daCunha2015PO,Shai2015PRE} aims at attacking the nodes with inter-community edges that are crucial to maintain the connectivity among communities. Practically, the removal costs for different nodes are not the same, so attack strategies could  also be designed to minimize the total costs\cite{Ren2019PNAS}.

Given an $N$-node network, which is subject to node-attacks, there are $N!$ possible attack sequences in total. Thus, it is quite possible to have different or even opposite conclusions for network robustness depending on some topological issues. For example, it is observed that homogeneous networks are more robust than heterogeneous networks against random attacks, MDTA, and MBTA\cite{Lu2016PO}. Also, when the attack strategy aims at removing the three-level tree structures (including random, maximum- and minimum-degree nodes)\cite{Hao2016PA}, homogeneous networks are more robust than heterogeneous networks. However, if one aims at removing approximately the longest simple path from a network, then homogeneous networks are more vulnerable than heterogeneous networks\cite{Pu2015PA}. Moreover, for networks with special topological features, the efficiencies of different attack strategies are also different; for example, MDTA causes greater damages to local-world networks\cite{Li2003PA} with  larger local-world sizes, while networks with smaller local-world sizes show better robustness regarding both connectivity and controllability\cite{Sun2016PLA}.

\subsubsection{Damage-based Attack Strategies}\label{subsub:dmg}

The concept of ``damage''\cite{Wang2014PA} in network connectivity helps to evaluate and guide attacks. The damage of a specific node is quantified by the change of the LCC size, before and after attacking the node. Therefore, it is natural that an efficient greedy attack strategy can be formed by sequentially attacking the node whose removal or malfunctioning will cause the greatest damage to the network\cite{Wang2014PA}. With damage as the importance measure, the most destructive node-removal sequence can be searched by solving a combinatorial optimization problem, using genetic algorithm\cite{Lin2022SC}, memetic algorithm\cite{Yang2018PA}, or other advanced optimization tools.

Different from the damage of connectivity, the damage of controllability is defined based on the categorization of edges or nodes. An edge or node is critical if and only if its removal increases the number of needed DNs; otherwise, it is non-critical\cite{Liu2011N,Sun2019ICSRS,Lou2021CNSNS}. The damage of controllability helps to form effective attack strategies, where critical edges or nodes will be removed with the highest priority\cite{Sun2019ICSRS,Lou2021CNSNS}.

Damage-based attack strategies are intuitive and the maximal destruction is guaranteed for every single attack. However, they have two clear disadvantages: 1) the maximal destruction of a series of continuous attacks cannot be guaranteed; 2) the computational cost of calculating the damage is not negligible.

\subsubsection{Computational Intelligence-based Attack Strategies}
Searching for a desirable attack sequence from the large number of possible choices is an NP-hard combinatorial optimization problem\cite{Karp1972CCC,Braunstein2016PNAS}. Evolutionary algorithms have been applied to dealing with this problem, such as genetic algorithms\cite{Zhang2016CEC}, artificial bee colony algorithm\cite{Lozano2017IS}, Tabu search algorithm\cite{Deng2016PA,Qi2018CHAOS}, and other metaheuristic algorithms\cite{Ventresca2012COR,Li2020ESA}. Candidate attack sequences referred to as individuals form a population, which are evolved towards the optimal destruction of networks.
Moreover, machine learning techniques have been increasingly used to explore optimal attack strategies on large-scale networks. Ensemble learning is employed to estimate node importance, where node damage (see Subsection \ref{subsub:dmg}) is used for training the model, such that nodes with higher damages can be identified, thus an efficient attack strategy can be designed\cite{Li2019ISBSCI}.
The minimal set of critical nodes is identified using graph attention networks\cite{Velivckovic2017arXiv}, which is then used to effectively disintegrate a complex network\cite{Grassia2021NC}. Such an attack strategy can be successful based on deep reinforcement learning\cite{Fan2020NMI}. A sequential attack process can also be modeled by a Markov decision process, whereas deep reinforcement learning\cite{Mnih2015N,Sutton2018Book} can be used to find optimal attack sequences\cite{Yan2016TIFS,Tian2021CSF,Yan2022TNSE}.
Recently, a combination of convolutional neural networks (CNN) and graph neural networks (GNN)\cite{Kipf2016arXiv,Hamilton2017NIPS,Hamilton2020Book} has been used for measuring the node importance in virus spreading models\cite{Zhang2022Neur}.

The computational intelligence-based attack strategies require a non-negligible or even substantial amount of computational cost in the stages of robustness evaluation and model training. The difference is that evolutionary algorithm-based strategies aim at finding the most destructive attack sequence for the given networks, while machine learning-based strategies also pursue the generalizability to unknown data, for which greater computational cost is needed in the training stage.

\subsection{Robustness Performance Prediction}\label{sub:pre}

Evaluating \textit{a posteriori} measures by attack simulations is generally very time-consuming. In case that the exact robustness values are not required, approximated values can be estimated by either analytical or computational methods. In so doing, the time complexity is either constant for analytical methods\cite{Lou2023IJCAS} or increasing significantly slower than that of attack simulations for computational methods\cite{Lou2022TCYB}.

\subsubsection{Analytical Approximation}

Analytical approximations require full knowledge of the network structure and the applied attack strategy that can be well-modeled\cite{Sun2019ICSRS,Cai2021TSMC}, such as random attacks. Given the network adjacency matrix, the controllability configuration and critical edges can be found, so that the controllability curve under random edge-attacks can be approximated based on the uniformly-random decreasing process of critical edges\cite{Sun2019ICSRS}. This analytical method is applicable to approximating the controllability robustness\cite{Sun2019ICSRS}, as shown in Eq. (\ref{eq:nde}), and the reachability-based controllability robustness\cite{Sun2021TNSM}, as shown in Eq. (\ref{eq:freec}), under random edge-attacks.

For random-graph networks, based on the fact that the generation mechanism is essentially the same as the random edge-removal process from a fully-connected network in a reverse manner, a precise approximation can be designed. The given random-graph networks are classified as either ``dense'', ``median'', or ``sparse''. Then, the hybrid approximation method uses different prior knowledge to approximate the controllability curves\cite{Lou2023IJCAS}.
In comparison, the approach of \cite{Lou2023IJCAS} performs significantly better in predicting the controllability curves of random-graph networks under random edge-attacks; while its disadvantage is clear that it is applicable only to the above-mentioned scenario.

\subsubsection{Machine Learning-based Prediction}

Machine learning algorithms, such as linear regression, random forest, and neural networks, have been successfully applied to predicting the number of DNs under random or targeted edge-attack, such that the controllability curves can be fitted\cite{Dhiman2021MLN}.

During the network robustness optimization processes, calculating the exact robustness values may not be required for every generation. Therefore, fast estimation can be used to improve the efficiency. For example, in Refs.\cite{Wang2020TEVC,Wang2021TEVC}, three algorithms, including radial basis function\cite{Hardy1971JGR}, inverse distance weighting\cite{Zhou2005CEC} and least-squares\cite{Shepard1968ACMNC}, form a surrogate ensemble for estimating connectivity robustness values; attack simulations are intermittently performed for obtaining real robustness values, which are used for simultaneously evaluation and updating the surrogates. The computational time of optimization can be significantly reduced by using such a surrogate ensemble\cite{Wang2020TEVC,Wang2021TEVC}.

The CNN-based prediction approach treats complex network data as gray-scale images\cite{Lou2022TCYB}, thereby fast approximating the robustness performance against different attacks in an end-to-end manner.  Prior knowledge is useful for pre-processing and filtering, which is utilized to build an improved predictor\cite{Lou2022TNNLS}, showing lower prediction errors. A limitation of this straightforward approach is that it cannot deal with the situation where the network size is significantly different from the input size of the CNN\cite{Wu2022IJCNN}. Graph representation learning\cite{Kipf2016arXiv,Hamilton2017NIPS,Hamilton2020Book}, which is specifically designed for processing graph data, provides a solution to overcome this problem\cite{Lou2022TCYB2}. In graph representation learning, the raw complex network data ($N\times N$) are compressed and unified to lower-dimensional representations ($W\times U$, with $W<N$ and $U\ll N$); thus, not only the input size problem is solved but also the topological features can be better extracted and utilized.

\subsection{Robustness Optimization}\label{sub:opt}

It is crucial to understand the relationship between the network structure and its functionality robustness. Generally, dense homogeneous networks have better robustness than spare heterogeneous ones, regarding the network connectivity, controllability, and communication ability.  However, it is also possible that well-designed heterogeneous networks have better robustness than homogeneous networks\cite{Yan2016SR}. For general heterogeneous networks, it is known that onion-like structures that possess higher assortativity coefficients are robust against attacks\cite{Herrmann2011JSTAT,Schneider2011PNAS,Wu2011PRE,Tanizawa2012PRE,Hayashi2018SR}. Given suitable robustness measure(s) as the objective(s), network robustness can be optimized using evolutionary algorithms\cite{Gunasekara2018MOO,Liu2019ECCN}.

\begin{figure}[htbp]
	\centering
	\includegraphics[width=.5\linewidth]{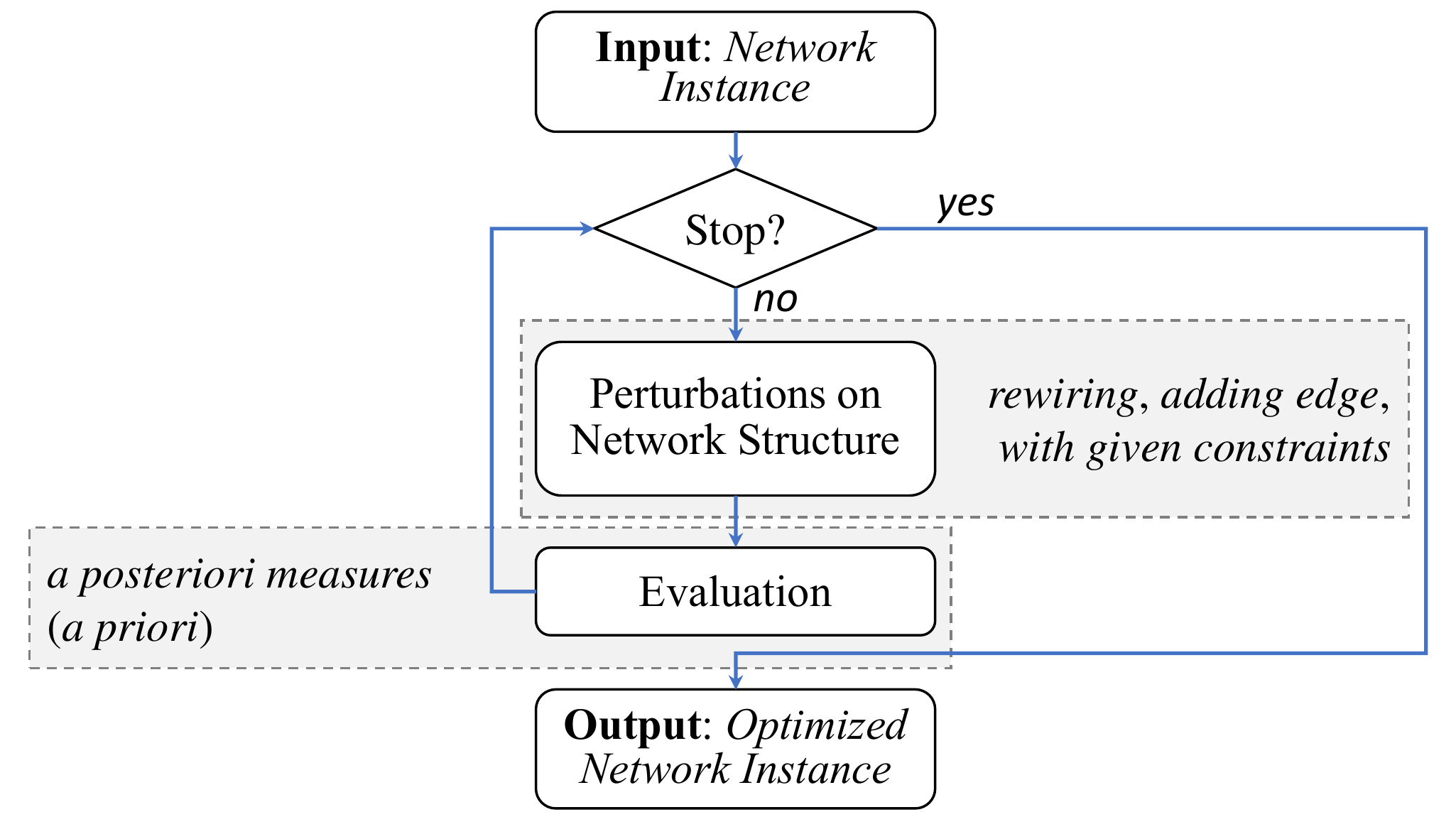}
	\caption{Flowchart of network robustness optimization using evolutionary algorithms. }\label{fig:ea}
\end{figure}

Figure \ref{fig:ea} shows a general flowchart of using evolutionary algorithms for network robustness optimization. Rewiring is the most widely-used strategy to perform disturbances onto the network structure, while in some specific applications adding edges is more cost-effective. Constraints such as degree preservation for all nodes guarantee some given prerequisites. After one or several edge rewiring operations, whether the disturbance enhances the robustness has to be evaluated by using either \textit{a priori} or \textit{a posteriori} measures. Here, within the focus of this survey, only the latter is discussed.

\begin{table*}[htbp]
	\centering \caption{Summary of using heuristic algorithms to optimize network robustness.}
	\begin{tabular}{|l|l|l|l|l|}
		\hline
		\multicolumn{1}{|c|}{Work} & \multicolumn{1}{c|}{Measure} & \multicolumn{1}{c|}{Algorithm} & \multicolumn{1}{c|}{Constraint} & \multicolumn{1}{c|}{\begin{tabular}[c]{@{}l@{}}Attack\\Object\end{tabular}} \\ \hline
		Schneider, et al.\cite{Schneider2011PNAS} & Eq. (\ref{eq:lc}) & random rewiring & \begin{tabular}[c]{@{}l@{}}degree preservation\\ for each node\end{tabular} & node \\ \hline
		Herrmann, et al.\cite{Herrmann2011JSTAT} & Eq. (\ref{eq:lc}) & Monte Carlo-based algorithm & \begin{tabular}[c]{@{}l@{}}degree distribution\\ preservation\end{tabular} & node \\ \hline
		Buesser, et. al.\cite{Buesser2011ICANCA} & Eq. (\ref{eq:lc}) & simulated annealing & \begin{tabular}[c]{@{}l@{}}degree preservation\\ for each node\end{tabular} & node \\ \hline
		Zeng and Liu\cite{Zeng2012PRE} & Eq. (\ref{eq:lc}) & hybrid greedy algorithm & \begin{tabular}[c]{@{}l@{}}degree preservation\\ for each node\end{tabular} & node and edge \\ \hline
		\begin{tabular}[c]{@{}l@{}}Peixoto and\\Bornholdt\cite{Peixoto2012PRL}\end{tabular} & Eq. (\ref{eq:lc}) & \begin{tabular}[c]{@{}l@{}}BFGS\cite{Fletcher2013Book} and\\ other swarm-based algorithms\end{tabular} & \begin{tabular}[c]{@{}l@{}} average degree\\preservation\end{tabular} & node \\ \hline
		Cao, et al.\cite{Cao2013CSF} & Eq. (\ref{eq:lc}) & strategies of adding edges & N/A & node \\ \hline
		Zhou and Liu\cite{Zhou2014PA} & Eq. (\ref{eq:lc}) & memetic algorithm & \begin{tabular}[c]{@{}l@{}}degree preservation\\ for each node\end{tabular} & node \\ \hline
		Xiao, et al.\cite{Xiao2014CPB} & Eq. (\ref{eq:xiao}) & dynamic optimization & \begin{tabular}[c]{@{}l@{}}degree preservation\\ of each node\end{tabular} & node \\ \hline
		Bai, et al.\cite{Bai2015CPL} & Eq. (\ref{eq:lc}) & hill-climbing search & \begin{tabular}[c]{@{}l@{}}degree preservation\\ for each node\end{tabular} & node \\ \hline
		Yang, et al.\cite{Yang2015PO} & Eq. (\ref{eq:lc}) & \begin{tabular}[c]{@{}l@{}}greedy bypass\\rewiring algorithm\end{tabular} & \begin{tabular}[c]{@{}l@{}}preserving both the\\ degree distribution and\\ community structure\end{tabular} & node \\ \hline
		Tang, et al.\cite{Tang2015EPL} & Eq. (\ref{eq:tang}) & memetic algorithm & \begin{tabular}[c]{@{}l@{}}degree preservation\\ for each node\end{tabular} & node \\ \hline
		Ma, et al.\cite{Ma2016PA} & Eq. (\ref{eq:lc}) and Eq. (\ref{eq:lce}) & edge-replenishment strategy & \begin{tabular}[c]{@{}l@{}}keep the total numbers\\ of nodes and edges\end{tabular} & node and edge \\ \hline
		Sun, et al.\cite{Sun2016PA} & Eq. (\ref{eq:lc}) & tabu search & \begin{tabular}[c]{@{}l@{}}degree preservation\\ for each node\end{tabular} & node \\ \hline
		Tang, et al.\cite{Tang2016SR} & Eq. (\ref{eq:tang}) & memetic algorithm & \begin{tabular}[c]{@{}l@{}}degree preservation\\ for each node\end{tabular} & node \\ \hline
		Park and Hahn\cite{Park2016PRE} & Eq. (\ref{eq:lc}) & \begin{tabular}[c]{@{}l@{}}greedy bypass\\ rewiring algorithm\end{tabular} & N/A & node \\ \hline
		Wang and Liu\cite{Wang2017SR} & \begin{tabular}[c]{@{}l@{}}2 objectives: \\ Eq. (\ref{eq:lc}) and cooperation\\ (fraction of cooperators)\end{tabular} & \begin{tabular}[c]{@{}l@{}}multi-objective\\ evolutionary algorithm\end{tabular} & \begin{tabular}[c]{@{}l@{}}degree preservation\\ for each node\end{tabular} & node \\ \hline
		Wang and Liu\cite{Wang2017CEC} & community integrity\cite{Ma2013PRE} & genetic algorithm & \begin{tabular}[c]{@{}l@{}}degree distribution \\ preservation\end{tabular} & node \\ \hline
		Wang, et al.\cite{Wang2017JSTAT} & \begin{tabular}[c]{@{}l@{}}normalized mutual \\ information\cite{Wang2017JSTAT}\end{tabular} & simulated annealing & \begin{tabular}[c]{@{}l@{}}degree distribution \\ preservation\end{tabular}& node \\ \hline
		Wang and Liu\cite{Wang2018TNSE} & \begin{tabular}[c]{@{}l@{}}2 objectives:\\ Eq. (\ref{eq:wang_tnse}) and cooperation\\ (fraction of cooperators)\end{tabular} & \begin{tabular}[c]{@{}l@{}}multi-objective\\ evolutionary algorithm\end{tabular} & \begin{tabular}[c]{@{}l@{}}degree preservation\\ for each node\end{tabular} & node \\ \hline
		Rong and Liu\cite{Rong2018PA} & Eq. (\ref{eq:lc}) & heuristic algorithm & \begin{tabular}[c]{@{}l@{}}degree preservation\\ for each node\end{tabular} & node \\ \hline
		Gunasekara, et al.\cite{Gunasekara2018MOO} &
		\begin{tabular}[c]{@{}l@{}}Eq. (\ref{eq:lc}) and two\\spectral measures\end{tabular}
		 & \begin{tabular}[c]{@{}l@{}}multi-objective \\ evolutionary algorithm\end{tabular} & \begin{tabular}[c]{@{}l@{}}N/A\end{tabular} & node \\ \hline
		Liu, et al.\cite{Liu2019TEVC} & Eq. (\ref{eq:lc}) & evolutionary algorithm & N/A & node \\ \hline
		Liu, et al.\cite{Liu2019ECCN} &
		\begin{tabular}[c]{@{}l@{}}Eq. (\ref{eq:lc}) and Eq. (\ref{eq:lce})\end{tabular}
		& \begin{tabular}[c]{@{}l@{}}multi-objective \\ evolutionary algorithm\end{tabular} & \begin{tabular}[c]{@{}l@{}}degree preservation\\ for each node\end{tabular} & node and edge \\ \hline
		Qiu, et al.\cite{Qiu2019TN} & Eq. (\ref{eq:lc}) & multi-population co-evolution & \begin{tabular}[c]{@{}l@{}}degree preservation\\ for each node\end{tabular} & node \\ \hline
		Cai, et al.\cite{Cai2020SSCI} & \begin{tabular}[c]{@{}l@{}}2 objectives:\\ 1) maximize algebraic\\ connectivity\\ 2) minimize the\\ number of removed edges\end{tabular} & \begin{tabular}[c]{@{}l@{}}NSGA-II, NSGA-III,\\ and MODPSO\end{tabular} & N/A & edge \\ \hline
		Wang, et al.\cite{Wang2020TEVC} & Eq. (\ref{eq:lc})* & \begin{tabular}[c]{@{}l@{}}surrogate-assisted\\ evolutionary algorithm\end{tabular} & \begin{tabular}[c]{@{}l@{}}degree preservation\\ for each node\end{tabular} & node \\ \hline
		Wang, et al.\cite{Wang2021TEVC} & Eq. (\ref{eq:lc})* & \begin{tabular}[c]{@{}l@{}}surrogate-assisted\\ multi-objective \\ evolutionary algorithm\end{tabular} & \begin{tabular}[c]{@{}l@{}}degree preservation\\ for each node\end{tabular} & node and edge \\ \hline
		Lou, et al.\cite{Lou2020TCASI} & Eq. (\ref{eq:nd}) & \begin{tabular}[c]{@{}l@{}}random edge rectification\end{tabular} & \begin{tabular}[c]{@{}l@{}} average degree\\preservation\end{tabular} & node and edge \\ \hline
		Lou, et al.\cite{Lou2021TCASII} & Eq. (\ref{eq:nd}) & \begin{tabular}[c]{@{}l@{}}random rewiring\end{tabular} & \begin{tabular}[c]{@{}l@{}} average degree\\preservation; and\\ underlying-topology\\preservation \end{tabular} & node \\ \hline
		\multicolumn{5}{l}{* with the assistance of surrogates and assortativity}
	\end{tabular}\label{tab:ea}
\end{table*}

Table \ref{tab:ea} summarizes 29 methods on network robustness optimization. The most common scenario of these methods is to use evolutionary algorithms to optimize network robustness under node-attacks measured by Eq. (\ref{eq:lc}), where the degree (or both in- and out-degrees) for each node remains unchanged during the optimization process. Extensions of this common scenario include considering different measures, using advanced algorithms, imposing different constraints of topology disturbances, and targeting different objects. Since a single measure sometimes cannot fully reflect the network robustness\cite{Liu2017FCS,Mburano2021ISNCC}, multiple robustness measures are usually considered, which are simultaneously optimized by using multi-objective optimization algorithms\cite{Gunasekara2018MOO,Liu2019ECCN,Wang2021TEVC}.

\section{Measuring Network Destruction}\label{sec:des}

It is observed that \textit{a posteriori} measures not only have intuitively clear meanings for a network function, as discussed in Subsection \ref{sub:fun}, but also have clearer descriptions about the network robustness, as introduced in Section \ref{sec:cmp}. One significant disadvantage of \textit{a posteriori} measures, however, is that their calculations are generally time-consuming. Nevertheless, in some applications, this can be (partially) solved by using analytical and computational methods, as discussed in Subsection \ref{sub:pre}.

Here, another common shortcoming of \textit{a posteriori} measures is addressed. The calculation of most \textit{a posteriori} measures is based on the entire process from attacking the first object to ending the attack at the last object. Practically, if a network has been severely destructed or malfunctioned, measuring its functionality will have no meaning. Also, complete disconnection of networks may not always be important in many applications. for example, when cascading failures are concerned, as shown in Eq. (\ref{eq:tang}), complete disconnection is unnecessary to attempt. Therefore, it is not always necessary to attack all nodes or edges for measuring the network robustness.

Clearly, determining when to stop attacking or whether a network is severely destructed or malfunctioned is application-dependent. For example, the robustness of food webs is widely measured by $R50$, which is the proportion of species (nodes) that has to be removed to cause the extinction of $50\%$ of the species in the food web\cite{Dunne2004MEPS,Curtsdotter2011BAE,Bellingeri2013TE}.

Different from the Molloy--Reed criterion\cite{Molloy1995RSA}, which states that a network will lose its giant component if $\langle k^2\rangle/\langle k\rangle >2$ is reached, here a new measure of network destruction is proposed, based on the change of the number of connected components (NCC)\cite{Uehara1999TR}. Specifically, for the \textit{a posteriori} measure $f(\cdot)$ that considers the network destruction, Eq. (\ref{eq:r}) can be rewritten as
\begin{equation}\label{eq:rt}
	R_{14}=\frac{1}{T+1}\sum_{i=0}^{T}{w_i\cdot f(i)}\,,
\end{equation}
where $T$ ($T<N$) is the counted number of removed objects before the threshold of ``severe destruction'' is reached. Here, the threshold integer $T$ separates the attack process into two parts: before $T$ is reached, the network is considered as normal; after $T$ is reached, the network is deemed breakdown.
Network robustness will be measured only before this threshold is reached. In the literature about node-attacks, $T$ is set to be $0.5N$ in\cite{Fan2020NMI}, $0.05N$ in\cite{Lordan2019RESS}, or less than $0.2N$ in\cite{Lu2016PO}. All are user-defined fixed integers.

In this paper, instead of setting $T$ to be a fixed value, the network destruction is considered from the perspective of NCC, which changes non-monotonically during the attack process.  In general, there is a clear turning point in the curve of NCC. For a connected network, its initial NCC is 1. The value of NCC increases as nodes and edges are being attacked. During the targeted attack process, the isolated nodes (generated due to attacks) will never be removed until there are only isolated nodes left in the residual network, since connected nodes are always targets if they exist. Therefore, the turning point of the NCC tendency curve can be used as the indicator of severe network destruction, namely, when this turning point appears, it means that there are only isolated nodes left in the residual network. This indicator of destruction is studied from a general perspective but not for a specific application. Note that removing isolated nodes is possible in any step of random attacks; hence, it is not suitable to use this turning point as the indicator of destruction.

The number of DN is non-decreasing and the numbers of LCC and CNP are non-increasing. Moreover, stagnation of DN, LCC, and CNP may occur frequently; therefore, it is difficult if not impossible to determine the network destruction using the changes of DN, LCC, or CNP.

Let $D(i)$ denote the NCC values after a total of $i$ nodes have been attacked, where $D(i)\in[1,N]$ and $i\in[0,N-1]$. 
The turning point of $D(i)$ is calculated by
\begin{equation}\label{eq:t}
	T=\underset{i}{\argmax}~{D(i)}\,.
\end{equation}
In attack simulations, $T$ can be determined when $D(i)<D(i-1)$ is successively detected for $\floor{pN}$ times. Then, $T=i-\floor{pN}$, where $p$ is a small decimal. It is empirically observed that the determination of $T$ is insensitive to the change of $p$. Set $p=0.05$ in the simulation, which means that when $D(i)<D(i-1)$ is successively detected for $\floor{0.05N}$ times, one has $T=i-\floor{0.05N}$.

Equation (\ref{eq:t}) can also be applied to edge-attacks. Since NCC will not decrease under edge-attacks, $T=i-\floor{pN}$ can be determined when $D(i)=D(i-1)$ is successively detected for $\floor{pN}$ times.

\begin{figure*}[htbp]
	\centering
	\includegraphics[width=\linewidth]{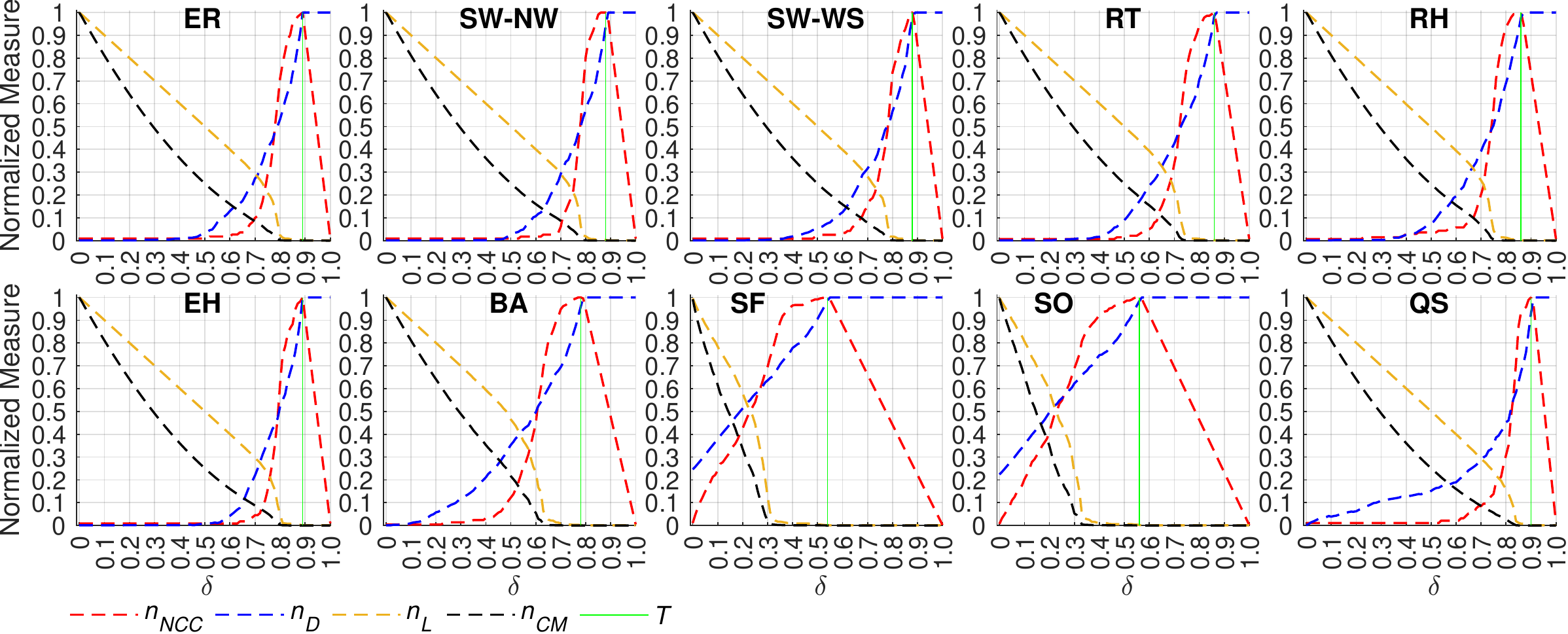}
	\caption{[color online] Network robustness in terms of the number of connected components ($n_{NCC}$), controllability ($n_D$), connectivity ($n_L$), and communication ability ($n_{CM}$), together with the threshold of destruction $T$. Here, $\delta$ represents the portion of nodes removed from the network. Node MDTA is implemented.}\label{fig:ndeg}
\end{figure*}

\begin{figure*}[htbp]
	\centering
	\includegraphics[width=\linewidth]{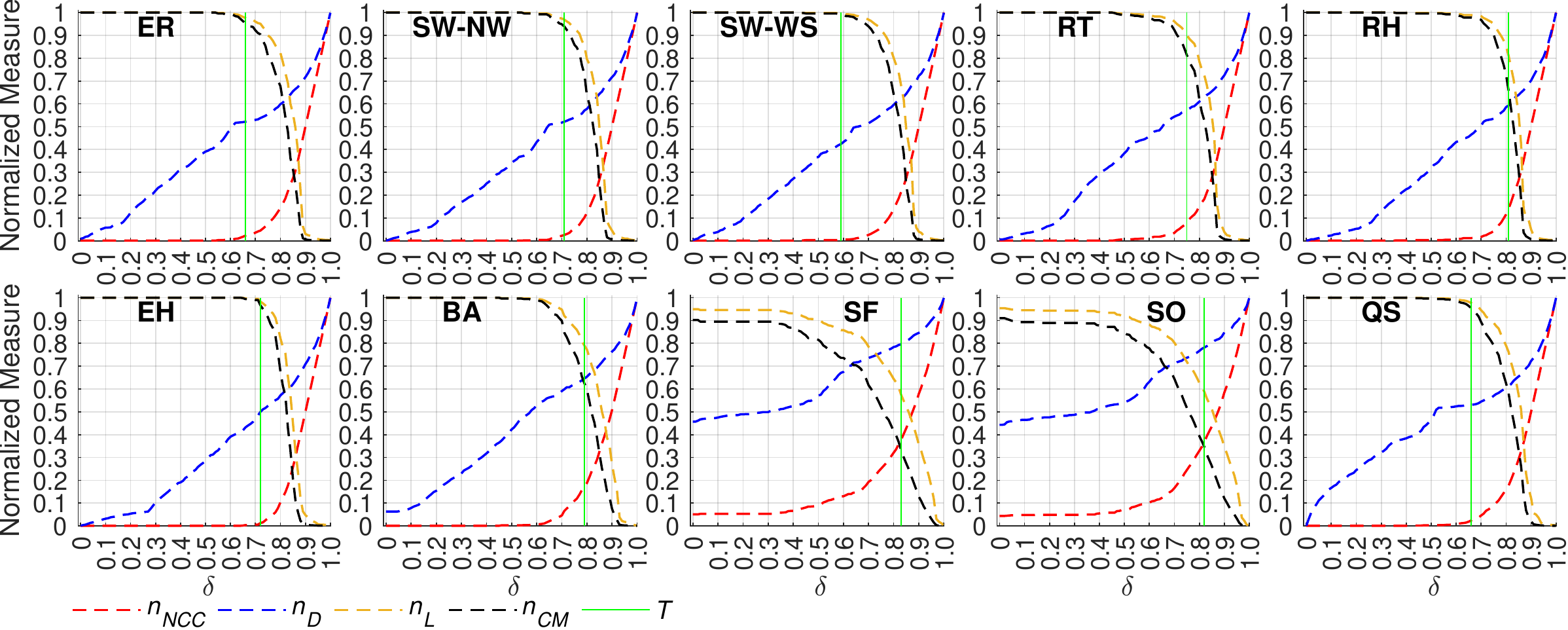}
	\caption{[color online][color online] Network robustness in terms of the number of connected components ($n_{NCC}$), controllability ($n_D$), connectivity ($n_L$), and communication ability ($n_{CM}$), together with the threshold of destruction $T$. Here, $\delta$ represents the portion of edges removed from the network. Edge MBTA is implemented. }\label{fig:ebet}
\end{figure*}

Figures \ref{fig:ndeg} and \ref{fig:ebet} show the attack simulation results under node MDTA and edge MBTA, respectively.

Here, a total of 10 synthetic network models are simulated, including the Erd{\"{o}}s--R{\'e}nyi (ER) random-graph\cite{Erdos1964RG}, Newman--Watts small-world (SW-NW)\cite{Newman1999PLA}, Watts--Strogatz small-world (SW-WS)\cite{Watts1998N}, random triangle (RT)\cite{Chen2019TCASII}, random hexagon (RH)\cite{Lou2022ACCESS}, extremely homogeneous (EH)\cite{Lou2020TCASI}, Barab{\'a}si--Albert (BA) scale-free\cite{Barabasi1999SCI,Barabasi2009SCI}, generic scale-free (SF)\cite{Goh2001PRL}, onion-like generic scale-free (OS)\cite{Schneider2011PNAS}, \textit{q}-snapback (QS)\cite{Lou2018TCASI,Wu2022TCNS} networks. In all simulations, the network size is set to $N=1000$ with $\langle k\rangle=10$. Four network functions are measured, that is, controllability robustness, connectivity robustness, communication robustness, and the number of connected components. The resultant values are normalized, so they are all in $[0,1]$.

As can be seen from Fig. \ref{fig:ndeg}, all the vertical green lines well match the turning points of the controllability curves (blue dashed curves). In contrast, for the connectivity and communication curves (brown and black dashed curves), the vertical green lines appear consistently later than the turning points of these curves.
This means that, under node MDTA, Eq. (\ref{eq:t}) suggests a good threshold for the network destruction in terms of controllability robustness, but not for connectivity and communication robustness. In contrast, as shown in Fig. \ref{fig:ebet}, the vertical green lines match the turning points of the connectivity and communication curves better than the controllability curves. This implies that, under edge MBTA, Eq. (\ref{eq:t}) suggests a good threshold for connectivity and communication robustness, but not for the controllability robustness.  This means that there is no turning point in the controllability curve under edge MBTA; the turning point only appears under node MDTA.

\begin{table*}[htbp]
	\centering \caption{Comparison of robustness performance under two measure schemes: complete disconnection (CD) and threshold-based disconnection (TD). Numbers in parentheses represent the corresponding ranks of robustness. }
	\begin{tabular}{|cc|llll|llll|}
		\hline
		\multicolumn{2}{|c|}{\multirow{2}{*}{}} & \multicolumn{4}{c|}{Node MDTA} & \multicolumn{4}{c|}{Edge MBTA} \\ \cline{3-10}
		\multicolumn{2}{|c|}{} & \multicolumn{1}{c|}{\begin{tabular}[c]{@{}l@{}}Controllability\\Robustness\end{tabular}} & \multicolumn{1}{c|}{\begin{tabular}[c]{@{}l@{}}Connectivity\\Robustness\end{tabular}} & \multicolumn{1}{c|}{\begin{tabular}[c]{@{}l@{}}Communication\\Robustness\end{tabular}} & \multicolumn{1}{c|}{$T$} & \multicolumn{1}{c|}{\begin{tabular}[c]{@{}l@{}}Controllability\\Robustness\end{tabular}} & \multicolumn{1}{c|}{\begin{tabular}[c]{@{}l@{}}Connectivity\\Robustness\end{tabular}} & \multicolumn{1}{c|}{\begin{tabular}[c]{@{}l@{}}Communication\\Robustness\end{tabular}} & \multicolumn{1}{c|}{$T$} \\ \hline
		\multicolumn{1}{|c|}{ER} & \begin{tabular}[c]{@{}c@{}}CD\\TD\end{tabular} & \multicolumn{1}{l|}{\begin{tabular}[c]{@{}l@{}}0.247 (2)\\0.155 (3)\end{tabular}} & \multicolumn{1}{l|}{\begin{tabular}[c]{@{}l@{}}0.476 (3)\\0.534 (6.5)\end{tabular}} & \multicolumn{1}{l|}{\begin{tabular}[c]{@{}l@{}}0.331 (3.5)\\0.372 (6.5)\end{tabular}} & 891 & \multicolumn{1}{l|}{\begin{tabular}[c]{@{}l@{}}0.390 (6)\\0.286 (4)\end{tabular}} & \multicolumn{1}{l|}{\begin{tabular}[c]{@{}l@{}}0.835 (4.5)\\0.991 (2)\end{tabular}} & \multicolumn{1}{l|}{\begin{tabular}[c]{@{}l@{}}0.806 (4.5)\\0.983 (2)\end{tabular}} & 381 \\ \hline
		\multicolumn{1}{|c|}{\begin{tabular}[c]{@{}c@{}}SW-\\NW\end{tabular}} & \begin{tabular}[c]{@{}c@{}}CD\\TD\end{tabular} & \multicolumn{1}{l|}{\begin{tabular}[c]{@{}l@{}}0.248 (3)\\0.146 (2)\end{tabular}} & \multicolumn{1}{l|}{\begin{tabular}[c]{@{}l@{}}0.474 (4)\\0.538 (1.5)\end{tabular}} & \multicolumn{1}{l|}{\begin{tabular}[c]{@{}l@{}}0.331 (3.5)\\0.376 (4)\end{tabular}} & 881 & \multicolumn{1}{l|}{\begin{tabular}[c]{@{}l@{}}0.361 (3)\\0.269 (3)\end{tabular}} & \multicolumn{1}{l|}{\begin{tabular}[c]{@{}l@{}}0.834 (6.5)\\0.985 (5)\end{tabular}} & \multicolumn{1}{l|}{\begin{tabular}[c]{@{}l@{}}0.808 (3)\\0.971 (5)\end{tabular}} & 406 \\ \hline
		\multicolumn{1}{|c|}{\begin{tabular}[c]{@{}c@{}}SW-\\WS\end{tabular}} & \begin{tabular}[c]{@{}c@{}}CD\\TD\end{tabular} & \multicolumn{1}{l|}{\begin{tabular}[c]{@{}l@{}}0.259 (4)\\0.159 (4)\end{tabular}} & \multicolumn{1}{l|}{\begin{tabular}[c]{@{}l@{}}0.473 (5)\\0.537 (3.5)\end{tabular}} & \multicolumn{1}{l|}{\begin{tabular}[c]{@{}l@{}}0.330 (5)\\0.375 (5)\end{tabular}} & 881 & \multicolumn{1}{l|}{\begin{tabular}[c]{@{}l@{}}0.372 (4)\\0.234 (1.5)\end{tabular}} & \multicolumn{1}{l|}{\begin{tabular}[c]{@{}l@{}}0.834 (6.5)\\0.998 (1)\end{tabular}} & \multicolumn{1}{l|}{\begin{tabular}[c]{@{}l@{}}0.806 (4.5)\\0.995 (1)\end{tabular}} & 347 \\ \hline
		\multicolumn{1}{|c|}{RT} & \begin{tabular}[c]{@{}c@{}}CD\\TD\end{tabular} & \multicolumn{1}{l|}{\begin{tabular}[c]{@{}l@{}}0.302 (7)\\0.190 (6)\end{tabular}} & \multicolumn{1}{l|}{\begin{tabular}[c]{@{}l@{}}0.459 (7)\\0.533 (8)\end{tabular}} & \multicolumn{1}{l|}{\begin{tabular}[c]{@{}l@{}}0.326 (7)\\0.378 (3)\end{tabular}} & 861 & \multicolumn{1}{l|}{\begin{tabular}[c]{@{}l@{}}0.376 (5)\\0.302 (6)\end{tabular}} & \multicolumn{1}{l|}{\begin{tabular}[c]{@{}l@{}}0.830 (8)\\0.964 (6)\end{tabular}} & \multicolumn{1}{l|}{\begin{tabular}[c]{@{}l@{}}0.801 (6.5)\\0.936 (6)\end{tabular}} & 426 \\ \hline
		\multicolumn{1}{|c|}{RH} & \begin{tabular}[c]{@{}c@{}}CD\\TD\end{tabular} & \multicolumn{1}{l|}{\begin{tabular}[c]{@{}l@{}}0.292 (6)\\0.178 (5)\end{tabular}} & \multicolumn{1}{l|}{\begin{tabular}[c]{@{}l@{}}0.464 (6)\\0.538 (1.5)\end{tabular}} & \multicolumn{1}{l|}{\begin{tabular}[c]{@{}l@{}}0.327 (6)\\0.379 (2)\end{tabular}} & 861 & \multicolumn{1}{l|}{\begin{tabular}[c]{@{}l@{}}0.344 (2)\\0.294 (5)\end{tabular}} & \multicolumn{1}{l|}{\begin{tabular}[c]{@{}l@{}}0.836 (2.5)\\0.918 (8)\end{tabular}} & \multicolumn{1}{l|}{\begin{tabular}[c]{@{}l@{}}0.812 (2)\\0.892 (8)\end{tabular}} & 455 \\ \hline
		\multicolumn{1}{|c|}{EH} & \begin{tabular}[c]{@{}c@{}}CD\\TD\end{tabular} & \multicolumn{1}{l|}{\begin{tabular}[c]{@{}l@{}}0.227 (1)\\0.132 (1)\end{tabular}} & \multicolumn{1}{l|}{\begin{tabular}[c]{@{}l@{}}0.478 (2)\\0.536 (5)\end{tabular}} & \multicolumn{1}{l|}{\begin{tabular}[c]{@{}l@{}}0.332 (2)\\0.372 (6.5)\end{tabular}} & 891 & \multicolumn{1}{l|}{\begin{tabular}[c]{@{}l@{}}0.327 (1)\\0.234 (1.5)\end{tabular}} & \multicolumn{1}{l|}{\begin{tabular}[c]{@{}l@{}}0.836 (2.5)\\0.986 (4)\end{tabular}} & \multicolumn{1}{l|}{\begin{tabular}[c]{@{}l@{}}0.813 (1)\\0.975 (4)\end{tabular}} & 411 \\ \hline
		\multicolumn{1}{|c|}{BA} & \begin{tabular}[c]{@{}c@{}}CD\\TD\end{tabular} & \multicolumn{1}{l|}{\begin{tabular}[c]{@{}l@{}}0.443 (8)\\0.288 (8)\end{tabular}} & \multicolumn{1}{l|}{\begin{tabular}[c]{@{}l@{}}0.418 (8)\\0.534 (6.5)\end{tabular}} & \multicolumn{1}{l|}{\begin{tabular}[c]{@{}l@{}}0.307 (8)\\0.393 (1)\end{tabular}} & 782 & \multicolumn{1}{l|}{\begin{tabular}[c]{@{}l@{}}0.424 (7)\\0.372 (8)\end{tabular}} & \multicolumn{1}{l|}{\begin{tabular}[c]{@{}l@{}}0.841 (1)\\0.936 (7)\end{tabular}} & \multicolumn{1}{l|}{\begin{tabular}[c]{@{}l@{}}0.800 (8)\\0.896 (7)\end{tabular}} & 446 \\ \hline
		\multicolumn{1}{|c|}{SF} & \begin{tabular}[c]{@{}c@{}}CD\\TD\end{tabular} & \multicolumn{1}{l|}{\begin{tabular}[c]{@{}l@{}}0.783 (10)\\0.601 (10)\end{tabular}} & \multicolumn{1}{l|}{\begin{tabular}[c]{@{}l@{}}0.205 (10)\\0.376 (10)\end{tabular}} & \multicolumn{1}{l|}{\begin{tabular}[c]{@{}l@{}}0.149 (10)\\0.273 (10)\end{tabular}} & 545 & \multicolumn{1}{l|}{\begin{tabular}[c]{@{}l@{}}0.632 (10)\\0.610 (10)\end{tabular}} & \multicolumn{1}{l|}{\begin{tabular}[c]{@{}l@{}}0.778 (10)\\0.829 (10)\end{tabular}} & \multicolumn{1}{l|}{\begin{tabular}[c]{@{}l@{}}0.669 (10)\\0.718 (10)\end{tabular}} & 465 \\ \hline
		\multicolumn{1}{|c|}{SO} & \begin{tabular}[c]{@{}c@{}}CD\\TD\end{tabular} & \multicolumn{1}{l|}{\begin{tabular}[c]{@{}l@{}}0.768 (9)\\0.589 (9)\end{tabular}} & \multicolumn{1}{l|}{\begin{tabular}[c]{@{}l@{}}0.215 (9)\\0.380 (9)\end{tabular}} & \multicolumn{1}{l|}{\begin{tabular}[c]{@{}l@{}}0.156 (9)\\0.276 (9)\end{tabular}} & 564 & \multicolumn{1}{l|}{\begin{tabular}[c]{@{}l@{}}0.615 (9)\\0.588 (9)\end{tabular}} & \multicolumn{1}{l|}{\begin{tabular}[c]{@{}l@{}}0.781 (9)\\0.839 (9)\end{tabular}} & \multicolumn{1}{l|}{\begin{tabular}[c]{@{}l@{}}0.676 (9)\\0.733 (9)\end{tabular}} & 460 \\ \hline
		\multicolumn{1}{|c|}{QS} & \begin{tabular}[c]{@{}c@{}}CD\\TD\end{tabular} & \multicolumn{1}{l|}{\begin{tabular}[c]{@{}l@{}}0.286 (5)\\0.207 (7)\end{tabular}} & \multicolumn{1}{l|}{\begin{tabular}[c]{@{}l@{}}0.484 (1)\\0.537 (3.5)\end{tabular}} & \multicolumn{1}{l|}{\begin{tabular}[c]{@{}l@{}}0.333 (1)\\0.369 (8)\end{tabular}} & 901 & \multicolumn{1}{l|}{\begin{tabular}[c]{@{}l@{}}0.448 (8)\\0.361 (7)\end{tabular}} & \multicolumn{1}{l|}{\begin{tabular}[c]{@{}l@{}}0.835 (4.5)\\0.989 (3)\end{tabular}} & \multicolumn{1}{l|}{\begin{tabular}[c]{@{}l@{}}0.801 (6.5)\\0.978 (3)\end{tabular}} & 381 \\ \hline
	\end{tabular}\label{tab:t}
\end{table*}

Table \ref{tab:t} shows the comparison of robustness performance under two measure schemes, namely the complete disconnection (CD) scheme as described by Eq. (\ref{eq:r}) and the threshold-based disconnection (TD) scheme as described by Eq. (\ref{eq:rt}). In the table, the numbers in parentheses represent the corresponding ranks of robustness. It is clear that, under different schemes, the robustness performance can be measured very differently.  The TD robustness measures are recommended (or even necessary) to use for the following reasons: 1) the resultant ranks are unique, such that the robustness measures can be distinguished for different networks; 2) the TD measures require much fewer numbers of attacks to measure the robustness and thus requires less computational time.

\section{Comparison between \textit{A Priori} and \textit{A Posteriori} Measures} \label{sec:cmp}

Now, experimental results on \textit{a priori} and \textit{a posteriori} measures are compared under 3 different node-attack strategies, namely exhaustive attack (EXA)\cite{Lou2020TCASI}, MDTA, and MBTA.

\begin{figure*}[htbp]
	\centering
	\includegraphics[width=\linewidth]{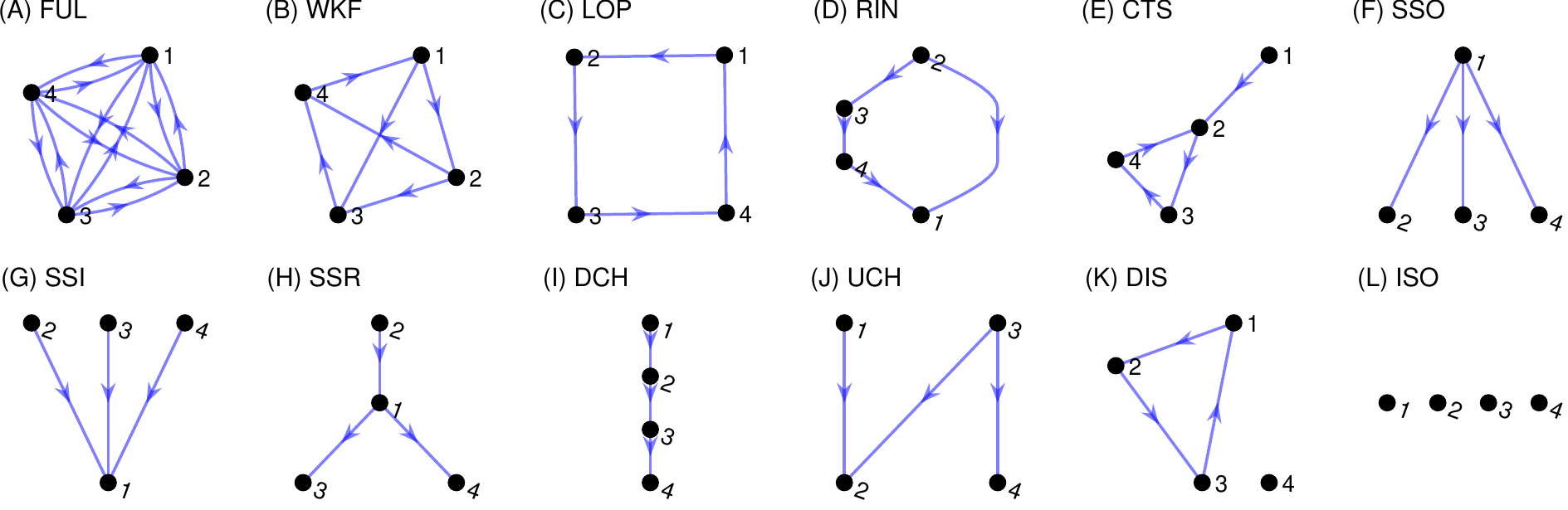}
	\caption{Twelve four-node directed networks: (A) fully-connected (FUL); (B) weak fully-connected (WKF); (C) loop (LOP); (D) ring-shaped non-loop (RIN); (E) cactus (CTS); (F) star-shaped out (SSO); (G) star-shaped in (SSI); (H) star-shaped random (SSR); (I) directed chain (DCH); (L) chain-shaped (UCH); (K) disconnected (DIS); and (L) isolated (ISO). }\label{fig:dir4}
\end{figure*}

\begin{figure*}[htbp]
	\centering
	\includegraphics[width=\linewidth]{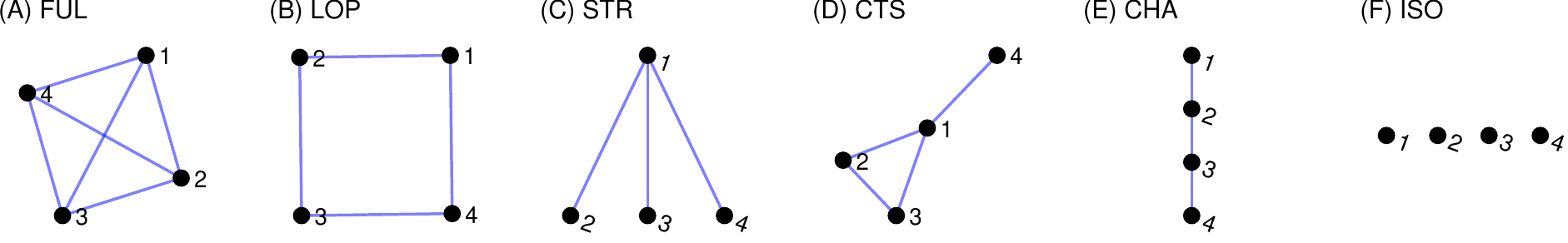}
	\caption{Six four-node undirected networks: (A) fully-connected (FUL); (B) loop (LOP); (C) star-shaped (STR); (D) cactus (CTS); (E) undirected chain (CHA); (F) isolated (ISO).  }\label{fig:undir4}
\end{figure*}

EXA averages the robustness values of a given $N$-node network over all the $N!$ possible node-attack sequences. Note that any intentional attack (e.g., MDTA, MBTA) is a particular case of the exhaustive attacks. Since the sample size of $N!$ becomes enormous as $N$ increases, only $N=4$ is tested here, which well serves for the purpose of demonstration. Figs. \ref{fig:dir4} and \ref{fig:undir4} show the topologies of the 4-node directed and undirected networks, respectively.


Four \textit{a posteriori} measures are simulated together, namely the connectivity robustness measured by Eq. (\ref{eq:lc}), controllability robustness measured by Eq. (\ref{eq:nd}), communication robustness measured by Eq. (\ref{eq:scom}), and connectivity robustness measured by NCC, which is defined as follows:
\begin{equation}\label{eq:ncc}
	R_{15}=\frac{1}{N}\sum_{i=0}^{N-1}{N_{NCC}(i)}\,,
\end{equation}
where $N_{NCC}(i)$ represents the number of connected components after a total of $i$ nodes have been attacked.

\subsection{A Priori Measures}\label{sub:pri}

\textit{A priori} measures are quantified by specific network features that can be calculated without performing attack simulations. \textit{A priori} measures require only one-time calculation and they usually have lower time and computational complexities compared to \textit{a posteriori} measures\cite{Chan2016DMKD,Freitas2022TKDE}.

Four topological \textit{a priori} measures are compared with \textit{a posteriori} measures for both directed and undirected networks, namely efficiency (EFF)\cite{Latora2001PRL}, node betweenness (NB)\cite{Freeman1977Soc}, edge betweenness (EB)\cite{Freeman1977Soc}, and clustering coefficient (CC)\cite{Watts1998N}.

Spectral measures are widely used to measure network connectivity robustness for undirected networks\cite{Chan2016DMKD}. Here, 3 adjacency matrix-based spectral measures, namely spectral radius (AS-SR)\cite{Van2011PRE}, spectral gap (AS-SG)\cite{Malliaros2012ICDM} and natural connectivity (AS-NC)\cite{Wu2010CPL} measures, and 3 Laplacian matrix-based spectral measures, namely algebraic connectivity (LS-AC)\cite{Fiedler1973CMJ}, number of spanning trees (LS-NS)\cite{Butler2008Book}, and effective resistance (LS-ER)\cite{Klein1993JMC}, are compared with \textit{a posteriori} measures for measuring undirected networks.

\subsection{Attack Simulations}
\begin{table*}[htbp]
	\centering \caption{Robustness ranks of the 12 directed 4-node networks, using 4 \textit{a posteriori} measures and 4 \textit{a priori} measures. }
	\begin{tabular}{|ccl|c|c|c|c|c|c|c|c|c|c|c|c|}
		\hline
		\multicolumn{3}{|c|}{\begin{tabular}[c]{@{}c@{}}Directed 4-Node\\ Networks\end{tabular}} & FUL & WKF & LOP & RIN & CTS & SSO & SSI & SSR & DCH & UCH & DIS & ISO \\\hline
		\multicolumn{1}{|c|}{\multirow{12}{*}{\rotatebox[origin=c]{90}{\textit{A Posteriori} Measures}}} & \multicolumn{1}{c|}{\multirow{4}{*}{EXA}} & $R_1$ (Eq.(\ref{eq:lc})) & 1.5 & 1.5 & 3.5 & 3.5 & 5 & 8 & 8 & 8 & 8 & 8 & 11 & 12 \\\cline{3-15}
		\multicolumn{1}{|c|}{} & \multicolumn{1}{c|}{} & $R_{15}$ (Eq.(\ref{eq:ncc})) & 1.5 & 1.5 & 3.5 & 3.5 & 5 & 8 & 8 & 8 & 8 & 8 & 11 & 12 \\\cline{3-15}
		\multicolumn{1}{|c|}{} & \multicolumn{1}{c|}{} & $R_3$ (Eq.(\ref{eq:nd})) & 1.5 & 1.5 & 3 & 4.5 & 4.5 & 10.5 & 10.5 & 8 & 6 & 9 & 7 & 12 \\\cline{3-15}
		\multicolumn{1}{|c|}{} & \multicolumn{1}{c|}{} & $R_7$ (Eq.(\ref{eq:scom})) & 1.5 & 1.5 & 3.5 & 3.5 & 5 & 7 & 7 & 7 & 9.5 & 9.5 & 11 & 12 \\\cline{2-15}
		\multicolumn{1}{|c|}{} & \multicolumn{1}{c|}{\multirow{4}{*}{MDTA}} & $R_1$ (Eq.(\ref{eq:lc})) & 3.5 & 3.5 & 7 & 3.5 & 3.5 & 10 & 3.5 & 10 & 3.5 & 8 & 10 & 12 \\\cline{3-15}
		\multicolumn{1}{|c|}{} & \multicolumn{1}{c|}{} & $R_{15}$ (Eq.(\ref{eq:ncc})) & 3.5 & 3.5 & 7 & 3.5 & 3.5 & 10 & 3.5 & 10 & 3.5 & 8 & 10 & 12 \\\cline{3-15}
		\multicolumn{1}{|c|}{} & \multicolumn{1}{c|}{} & $R_3$ (Eq.(\ref{eq:nd})) & 2.5 & 2.5 & 6 & 2.5 & 5 & 11 & 7.5 & 10 & 2.5 & 9 & 7.5 & 12 \\\cline{3-15}
		\multicolumn{1}{|c|}{} & \multicolumn{1}{c|}{} & $R_7$ (Eq.(\ref{eq:scom})) & 3.5 & 3.5 & 7 & 3.5 & 3.5 & 9.5 & 3.5 & 9.5 & 3.5 & 8 & 11 & 12 \\\cline{2-15}
		\multicolumn{1}{|c|}{} & \multicolumn{1}{c|}{\multirow{4}{*}{MBTA}} & $R_1$ (Eq.(\ref{eq:lc})) & 2.5 & 2.5 & 6.5 & 2.5 & 6.5 & 6.5 & 2.5 & 10.5 & 9 & 6.5 & 10.5 & 12 \\\cline{3-15}
		\multicolumn{1}{|c|}{} & \multicolumn{1}{c|}{} & $R_{15}$ (Eq.(\ref{eq:ncc})) & 2.5 & 2.5 & 6.5 & 2.5 & 6.5 & 6.5 & 2.5 & 10.5 & 9 & 6.5 & 10.5 & 12 \\\cline{3-15}
		\multicolumn{1}{|c|}{} & \multicolumn{1}{c|}{} & $R_3$ (Eq.(\ref{eq:nd})) & 1.5 & 1.5 & 5 & 3.5 & 3.5 & 10 & 7 & 11 & 7 & 9 & 7 & 12 \\\cline{3-15}
		\multicolumn{1}{|c|}{} & \multicolumn{1}{c|}{} & $R_7$ (Eq.(\ref{eq:scom})) & 2.5 & 2.5 & 6 & 2.5 & 8 & 6 & 2.5 & 10 & 9 & 6 & 11 & 12 \\\hline
		\multicolumn{2}{|c|}{\multirow{4}{*}{\textit{A Priori} Measures}} & EFF & 1 & 2 & 3 & 5 & 4 & 10 & 10 & 8 & 7 & 10 & 6 & 12 \\\cline{3-15}
		\multicolumn{2}{|c|}{} & NB & 2.5 & 10 & 11 & 5.5 & 9 & 2.5 & 2.5 & 5.5 & 8 & 2.5 & 7 & NA\\\cline{3-15}
		\multicolumn{2}{|c|}{} & EB & 2.5 & 8 & 11 & 5 & 10 & 2.5 & 2.5 & 6 & 9 & 2.5 & 7 & NA\\\cline{3-15}
		\multicolumn{2}{|c|}{} & CC & 1.5 & 1.5 & 8.5 & 8.5 & 4 & 8.5 & 8.5 & 8.5 & 8.5 & 8.5 & 3 & 8.5 \\\hline
	\end{tabular}\label{tab:dir}
\end{table*}

\begin{table}[htbp]
	\centering \caption{Robustness ranks of the 6 undirected 4-node networks, using 4 \textit{a posteriori} measures and 10 \textit{a priori} measures. }
	\begin{tabular}{|ccl|c|c|c|c|c|c|}
		\hline
		\multicolumn{3}{|c|}{\begin{tabular}[c]{@{}c@{}}Undirected 4-Node\\ Networks\end{tabular}} & FUL & LOP & STR & CTS & CHA & ISO \\\hline
		\multicolumn{1}{|c|}{\multirow{12}{*}{\rotatebox[origin=c]{90}{\textit{A Posteriori} Measures}}} & \multicolumn{1}{c|}{\multirow{4}{*}{\rotatebox[origin=c]{90}{EXA}}} & $R_1$ (Eq.(\ref{eq:lc})) & 1 & 2 & 4.5 & 3 & 4.5 & 6 \\\cline{3-9}
		\multicolumn{1}{|c|}{} & \multicolumn{1}{c|}{} & $R_{15}$ (Eq.(\ref{eq:ncc})) & 1 & 2 & 4.5 & 3 & 4.5 & 6 \\\cline{3-9}
		\multicolumn{1}{|c|}{} & \multicolumn{1}{c|}{} & $R_3$ (Eq.(\ref{eq:nd})) & 1 & 4 & 5 & 2 & 3 & 6 \\\cline{3-9}
		\multicolumn{1}{|c|}{} & \multicolumn{1}{c|}{} & $R_7$ (Eq.(\ref{eq:scom})) & 1 & 2 & 4 & 3 & 5 & 6 \\\cline{2-9}
		\multicolumn{1}{|c|}{} & \multicolumn{1}{c|}{\multirow{4}{*}{\rotatebox[origin=c]{90}{MDTA}}} & $R_1$ (Eq.(\ref{eq:lc})) & 1 & 2 & 5 & 3.5 & 3.5 & 6 \\\cline{3-9}
		\multicolumn{1}{|c|}{} & \multicolumn{1}{c|}{} & $R_{15}$ (Eq.(\ref{eq:ncc})) & 1 & 2 & 5 & 3.5 & 3.5 & 6 \\\cline{3-9}
		\multicolumn{1}{|c|}{} & \multicolumn{1}{c|}{} & $R_3$ (Eq.(\ref{eq:nd})) & 1 & 4 & 5 & 2.5 & 2.5 & 6 \\\cline{3-9}
		\multicolumn{1}{|c|}{} & \multicolumn{1}{c|}{} & $R_7$ (Eq.(\ref{eq:scom})) & 1 & 2 & 5 & 3.5 & 3.5 & 6 \\\cline{2-9}
		\multicolumn{1}{|c|}{} & \multicolumn{1}{c|}{\multirow{4}{*}{\rotatebox[origin=c]{90}{MBTA}}} & $R_1$ (Eq.(\ref{eq:lc})) & 1 & 2.5 & 5 & 2.5 & 4 & 6 \\\cline{3-9}
		\multicolumn{1}{|c|}{} & \multicolumn{1}{c|}{} & $R_{15}$ (Eq.(\ref{eq:ncc})) & 1 & 2.5 & 5 & 2.5 & 4 & 6 \\\cline{3-9}
		\multicolumn{1}{|c|}{} & \multicolumn{1}{c|}{} & $R_3$ (Eq.(\ref{eq:nd})) & 1.5 & 4 & 5 & 1.5 & 3 & 6 \\\cline{3-9}
		\multicolumn{1}{|c|}{} & \multicolumn{1}{c|}{} & $R_7$ (Eq.(\ref{eq:scom})) & 1 & 2 & 5 & 3 & 4 & 6 \\\hline
		\multicolumn{2}{|c|}{\multirow{10}{*}{\rotatebox[origin=c]{90}{\textit{A Priori} Measures}}} & EFF & 1 & 2.5 & 4 & 2.5 & 5 & 6 \\\cline{3-9}
		\multicolumn{2}{|c|}{} & NB & 1 & 2.5 & 4 & 2.5 & 5 & NA\\\cline{3-9}
		\multicolumn{2}{|c|}{} & EB & 1 & 2.5 & 4 & 2.5 & 5 & NA\\\cline{3-9}
		\multicolumn{2}{|c|}{} & CC & 1 & 4.5 & 4.5 & 2 & 4.5 & 4.5 \\\cline{3-9}
		\multicolumn{2}{|c|}{} & AS-SR & 1 & 3 & 4 & 2 & 5 & 6 \\\cline{3-9}
		\multicolumn{2}{|c|}{} & AS-SG & 1 & 2 & 4 & 3 & 5 & 6 \\\cline{3-9}
		\multicolumn{2}{|c|}{} & AS-NC & 1 & 3 & 4 & 2 & 5 & 6 \\\cline{3-9}
		\multicolumn{2}{|c|}{} & LS-AC & 1 & 2 & 3.5 & 3.5 & 5 & 6 \\\cline{3-9}
		\multicolumn{2}{|c|}{} & LS-NS & 1 & 2.5 & 4.5 & 2.5 & 4.5 & 6 \\\cline{3-9}
		\multicolumn{2}{|c|}{} & LS-ER & 1 & 2 & 4 & 3 & 5 & NA\\\hline
	\end{tabular}\label{tab:und}
\end{table}

Tables \ref{tab:dir} and \ref{tab:und} show the robustness ranks of different 4-node networks, where equal robustness share the same ranks. Intensive simulations show the following general results:

1) \textit{A posteriori} measures return different values when different attack strategies are applied, while \textit{a priori} measures return a unique value for the network under different attacks. This suggests that \textit{a priori} measures are not capable of distinguishing the network robustness under different attacks; while \textit{a posteriori} measures are capable of capturing even tiny robustness difference of a network under different attacks.

2) Most measures can reflect the basic fact that fully-connected networks possesses the best robustness, followed by connected networks, then disconnected networks, and lastly isolated networks, which possess the worst robustness. This means that both \textit{a posteriori} and \textit{a priori} measures can generally reflect the robustness of networks, regardless of the attack strategies.

3) For the 12 directed networks, no measure can distinguish all of them with different robustness values; whereas for the 6 undirected networks, the measures described by Eq. (\ref{eq:nd}) under EXA, Eq. (\ref{eq:scom}) under EXA and MBTA, as well as the 3 adjacency matrix-based spectral measures, all return 6 distinguished robustness values. This suggests that the robustness measures for directed networks remain a challenging research topic for future investigation. It is also intuitively clear that different directed networks may possess the same robustness performance, which are impossible to distinguish.

4) The two connectivity robustness measures, $R_1$ in Eq. (\ref{eq:lc}) and $R_{15}$ in Eq. (\ref{eq:ncc}), return identical robustness values (ranks) in all the cases for both directed and undirected networks, indicating that these two measures are highly-correlated.  It is worth mentioning that $R_1$ has been widely used for robustness measure, while $R_{15}$ is rarely used. As can be seen from Section \ref{sec:des}, more useful information can be dug out from NCC, which may be underestimated as compared to LCC.

5) CC cannot provide distinguished robustness values for different networks in many cases; while NB, EB, and LS-ER cannot measure the isolated networks. In contrast, the \textit{a posteriori} measures are able to return different robustness values for all networks. This suggests a wider applicability of the \textit{a posteriori} measures.

The advantages of \textit{a posteriori} measures, which are summarized from the experimental comparisons shown in Tables \ref{tab:dir} and \ref{tab:und}, are as follows:

1) \textit{A posteriori} measures have intuitively clear meanings for each network function and each attack strategy, while \textit{a priori} measures always return a unique robustness value for a given network.

2) \textit{A posteriori} measures can provide more robustness information about the given networks. For example, although the 3 adjacency-spectral measures can return distinguished robustness values for different undirected networks, as shown in Table \ref{tab:und}, they disagree with the assertion that LOP is more robustness than CTS.  In contrast, \textit{a posteriori} measures tell much more useful information: LOP is more robustness than CTS with respect to $R_1$, $R_{15}$, and $R_7$, but CTS is more robustness than LOP with respect to $R_3$ under EXA and MDTA; two networks are equally robust under MBTA with respect to $R_1$ and $R_{15}$.

3) It is clear that the spectral measures provide better indicators than the topological measures, but the former can only be applied to undirected networks. Moreover, NB, EB, and LS-ER cannot even be calculated for isolated networks.  In contrast, \textit{a posteriori} measures can be applied to any network, and provide better indicators to network robustness, thus have a wider applicability.

\section{Prospective Research Directions}\label{sec:ftw}

Some prospective research directions are summarized from four aspects: 1) exploring better weighting methods and termination criteria for Eq. (\ref{eq:r}); 2) designing more efficient and precise analytical and computational estimation methods; 3) performing more efficient robustness optimization; and 4) exploring more real-world applications.

\subsection{Weighting the Attacks}

The currently widely-used \textit{a posteriori} robustness measures assign unique weights for each single attack in the attack sequence, assuming equal contributions of all the remaining network functionalities to the calculation of the overall robustness. As shown in Eq. (\ref{eq:r}), when $w_i=1/N$, it means that $f(i)$ and $f(j)$ ($j\neq i$) are equally important to the overall network robustness. Although the importance of attacking nodes $i$ and $j$ can be partially reflected by the different values of $f(i)$ and $f(j)$, this is clearly insufficient in many scenarios.

Practically, the removal or malfunction of some nodes will cause greater damages than other nodes. If such important nodes are attacked at the very beginning, the robustness measure should be different from the scenario that these important nodes can be protected until the later stages. The robustness measure can be delicately adjusted for different applications, by setting proper configuration of $w_i$. Meanwhile, if important nodes can be attacked at the beginning, more credits should be assigned to the attack strategy; otherwise, it means that the attack strategy is less efficient.

Moreover, the network robustness under different attack strategies can be weighted, depending on the specific situation and concern. Weighting values can be added into Eq. (\ref{eq:rpt}), where its current form implies uniform weights for different attack strategies. In practice, if the probabilities of a network suffering different attacks are different, then it is meaningful to impose different weights to them.

Possible realistic weighting methodologies include decaying weights, importance-based weights, adaptive weights, etc.

\subsection{Termination Criteria}

A realistic threshold of destruction is introduced in Section \ref{sec:des}, which gives an alternative threshold to the conventional settings, such as the Molloy--Reed criterion\cite{Molloy1995RSA} and the fixed-proportion threshold. However, there still lacks a systematic investigation on the determination of the time when a networked system is deemed breakdown thereby the attack process can be terminated. The destruction of networks can be investigated from the perspectives of topological structures, network functions, or both.

To determine proper termination criteria, analytical and theoretical studies can be carried out, for example, further development of the Molloy--Reed criterion\cite{Molloy1995RSA}, percolation theory\cite{Li2021PR}, and so on.  Empirical studies such as the realistic threshold introduced in Section \ref{sec:des} can also be further investigated. Moreover, machine learning techniques may be utilized for solving this problem more effectively from a data-scientific perspective, based on both real-world networks and synthetic models. For example, given real-world data of network destruction as training data, machine learning can be used to estimate whether a given network is considered breakdown, or when it would be breakdown, under attacks.

\subsection{Robustness Estimation}\label{sub:est}

It is important to precisely and cost-efficiently approximate various robustness of large-scale networks.

The existing analytical approximations are applicable only to very limited specific issues of complex networks, e.g., controllability robustness under random or critical edge-attacks\cite{Sun2019ICSRS,Sun2021TNSM,Lou2023IJCAS}. Considering Eq. (\ref{eq:nd}) as the controllability robustness measure, attacking a single node (or edge) may either increase the number of DN by 1, or it does not change the number of DN at all. Thus, the maximum damage to the network controllability is limited. In contrast, when Eq. (\ref{eq:lc}) is used to study the connectivity robustness, the range of damages caused by each attack to LCC could vary from 0 to $N-1$, namely with all possibilities. Therefore, predicting the connectivity robustness is much more uncertain and challenging than predicting the controllability robustness, either analytically or computationally\cite{Lou2021TNSE}.

In this direction, if the pattern of malicious attacks can be well modeled using mathematics and statistics tools, then analytical approximation methods are recommended; but if there is no such a pattern (neither random not specifically targeted), then analytical methods are inapplicable while computational techniques are effective.

A comprehensive investigation of analytical approximation to robustness is needed, where some potential research topics include: 1) modeling more intrinsic attacks other than random or degree-based attacks; 2) exploring the relationships between the topological features and the robustness performance, where if direct relationships cannot be revealed then indirected relationships may be explored, for example some critical points (e.g., turning points) of the robustness curve might be estimated using topological features, so that a robustness curve can be fitted based on these critical points. As for computational approaches, not only the state-of-the-art machine learning techniques can be developed and applied, but also prior knowledge and theoretical findings can be used to further improve the prediction performances.

\subsection{Robustness Optimization}

Network robustness optimization via topological rewiring is NP-hard\cite{Kempe2003KDD}. The development of evolutionary algorithms helps in effectively resolving this difficult problem.
Robustness optimization for large-scale complex networks is higher-dimensional and computational expensive in general. In this regard, dimension reduction can be archived by applying graph embedding or using GNN\cite{Kipf2016arXiv,Hamilton2017NIPS,Hamilton2020Book}, which not only compress higher-dimensional network data into lower-dimensional representations, but also extract structural features for further processing.
As for the computational expenses in robustness evaluation, surrogate models are advantageous for improving the search efficiency and capability\cite{Wang2020TEVC,Wang2021TEVC}. Robustness estimation techniques, as introduced in Subsection \ref{sub:est}, can provide even better estimation tools than the commonly-used surrogates methods in evolutionary computation.

Since the thriving development of evolutionary computation has provided useful approaches for complex networks to evolve towards more robust structures, the key issue in this research direction is how to substantially reduce the computational cost of robustness evaluation. Although surrogates and easy-to-access indicators (such as assortativity coefficient) have been employed, the runtime of optimization remains high in many real-world applications\cite{Wang2020TEVC,Wang2021TEVC}.

Fast and precise estimation methods, both analytical and computational, can be applied to further reduce the runtime. Since real evaluations of robustness are inevitable, the ratio and arrangement between the real evaluations and the estimations should be investigated, such that the cost-efficiency can be maximized. Moreover, instead of using adjacency matrices as the chromosomes, better network representations may be explored, such that the feasibility of robustness optimization in the lower-dimensional representation domain (other than the higher-dimensional topological domain) can be explored.

\subsection{Real-world Applications}

The study of network robustness not only has been extended to many different network types, including weighted networks\cite{Bellingeri2018PA}, network of networks\cite{Gao2011PRL,Dong2013PRE,Havlin2014EPJST,Liu2015CSF}, interdependent networks\cite{Huang2011PRE,Gao2012PRE,Cui2018PA,Gao2018PA,Zhang2018PO}, and multiplex networks\cite{Min2014PRE,Chen2017TKDE}, but also has been applied to more and more real-world applications, for example land and air transport networks\cite{Zhang2013EPL,Sun2017CJA,Yang2018TITS,Lordan2014TRPE,Cai2020SSCI,Jiao2020TRPD,Li2020TRPA,Lordan2020PO,Zhu2018PA,De2012TRPC}, wireless sensor networks\cite{Qiu2017TN,Qiu2019TN,Hu2020CC}, power grids\cite{Guo2017TPS,Chen2017TCASII,Tu2018TCASII}, Internet of Things\cite{Qiu2022Book}, and so on.

Together with the development of realistic robustness measures, fast and precise robustness estimation, and cost-efficient optimization techniques, it is expected that these findings and the developed techniques can significantly extend and facilitate broader applications of real-world network problems in the near future.

\section{Conclusions}\label{sec:end}

The rapid development of complex networks research demands effective measures on various types of network robustness, especially for practical \textit{a posteriori} measures.

This survey presents a summary and overview of the comprehensive network robustness research development, focusing on the \textit{a posteriori} robustness measures.  Specifically, the \textit{a posteriori} robustness measures are reviewed from four perspectives, namely the network functionality, malicious attacks, robustness estimation, and network robustness optimization. Moreover, a practical threshold of network destruction due to attacks is introduced. Network robustness is suggested to be measured only before the threshold of destruction is reached, thereafter the network is deemed breakdown and so further measuring its functionality is not meaningful anymore. Extensive simulations confirm that the proposed threshold is suitable for certain functional robustness under some specific attack strategies. Thereby, further systematic investigation is recommended for determining network destruction with respect to \textit{a posteriori} measures.

Moreover, experimental comparisons of \textit{a posteriori} and \textit{a priori} measures on directed and undirected example networks are performed and analyzed. Compared to \textit{a priori} measures, the advantages of \textit{a posteriori} measures are obvious: 1) \textit{a posteriori} measures have intuitively clear meanings for every network function and attack strategy; 2) \textit{a posteriori} measures provide more useful robustness information; and 3) \textit{a posteriori} measures have wider applicability.

Finally, some prospective research directions with respect to \textit{a posteriori} robustness measures are suggested, including weighting and termination of \textit{a posteriori} measures, analytical and computation-based robustness estimation methods, robustness optimization techniques, and some potential real-world applications.

\bibliographystyle{IEEEtran}
\bibliography{ref}

\end{document}